\documentclass[preprint2]{aastex62}
\newcommand{\U}[1]{\ensuremath{\mathrm{\ #1}}}
\newcommand{\UU}[2]{\ensuremath{\mathrm{\ #1^{#2}}}}
\usepackage{graphicx}
\usepackage{gensymb}
\usepackage{lineno}
\usepackage{multirow}
\usepackage{booktabs}
\usepackage{tabularx}
\usepackage[flushleft]{threeparttable}
\usepackage{ragged2e} 
\usepackage{color}
\usepackage[normalem]{ulem}
\usepackage{bm}
\usepackage[utf8]{inputenc}
\shorttitle{VERITAS and \textit{Fermi}-LAT observations of new HAWC sources}
\shortauthors{VERITAS, \textit{Fermi}-LAT, and HAWC Collaboration}

\begin{document}

\title{VERITAS and \textit{Fermi}-LAT observations of  TeV gamma-ray sources discovered by HAWC in the 2HWC catalog}

\author{A.~U.~Abeysekara}
\affiliation{Department of Physics and Astronomy, University of Utah, Salt Lake City, UT 84112, USA}
\author{A.~Archer}
\affiliation{Department of Physics, Washington University, St. Louis, MO 63130, USA}
\author{W.~Benbow}
\affiliation{Fred Lawrence Whipple Observatory, Harvard-Smithsonian Center for Astrophysics, Amado, AZ 85645, USA}
\author{R.~Bird}
\affiliation{Department of Physics and Astronomy, University of California, Los Angeles, CA 90095, USA}
\author{R.~Brose}
\affiliation{Institute of Physics and Astronomy, University of Potsdam, 14476 Potsdam-Golm, Germany}
\affiliation{DESY, Platanenallee 6, 15738 Zeuthen, Germany}
\author{M.~Buchovecky}
\affiliation{Department of Physics, Washington University, St. Louis, MO 63130, USA}
\author{J.~H.~Buckley}
\affiliation{Department of Physics, Washington University, St. Louis, MO 63130, USA}
\author{V.~Bugaev}
\affiliation{Department of Physics, Washington University, St. Louis, MO 63130, USA}
\author{A.~J.~Chromey}
\affiliation{Department of Physics and Astronomy, Iowa State University, Ames, IA 50011, USA}
\author{M.~P.~Connolly}
\affiliation{School of Physics, National University of Ireland Galway, University Road, Galway, Ireland}
\author{W.~Cui}
\affiliation{Department of Physics and Astronomy, Purdue University, West Lafayette, IN 47907, USA}
\affiliation{Department of Physics and Center for Astrophysics, Tsinghua University, Beijing 100084, China}
\author{M.~K.~Daniel}
\affiliation{Fred Lawrence Whipple Observatory, Harvard-Smithsonian Center for Astrophysics, Amado, AZ 85645, USA}
\author{A.~Falcone}
\affiliation{Department of Astronomy and Astrophysics, 525 Davey Lab, Pennsylvania State University, University Park, PA 16802, USA}
\author{Q.~Feng}
\affiliation{Physics Department, McGill University, Montreal, QC H3A 2T8, Canada}
\author{J.~P.~Finley}
\affiliation{Department of Physics and Astronomy, Purdue University, West Lafayette, IN 47907, USA}
\author{L.~Fortson}
\affiliation{School of Physics and Astronomy, University of Minnesota, Minneapolis, MN 55455, USA}
\author{A.~Furniss}
\affiliation{Department of Physics, California State University - East Bay, Hayward, CA 94542, USA}
\author{M.~H\"utten}
\affiliation{DESY, Platanenallee 6, 15738 Zeuthen, Germany}
\author{D.~Hanna}
\affiliation{Physics Department, McGill University, Montreal, QC H3A 2T8, Canada}
\author{O.~Hervet}
\affiliation{Santa Cruz Institute for Particle Physics and Department of Physics, University of California, Santa Cruz, CA 95064, USA}
\author{J.~Holder}
\affiliation{Department of Physics and Astronomy and the Bartol Research Institute, University of Delaware, Newark, DE 19716, USA}
\author{G.~Hughes}
\affiliation{Fred Lawrence Whipple Observatory, Harvard-Smithsonian Center for Astrophysics, Amado, AZ 85645, USA}
\author{T.~B.~Humensky}
\affiliation{Physics Department, Columbia University, New York, NY 10027, USA}
\author{C.~A.~Johnson}
\affiliation{Santa Cruz Institute for Particle Physics and Department of Physics, University of California, Santa Cruz, CA 95064, USA}
\author{P.~Kaaret}
\affiliation{Department of Physics and Astronomy, University of Iowa, Van Allen Hall, Iowa City, IA 52242, USA}
\author{P.~Kar}
\affiliation{Department of Physics and Astronomy, University of Utah, Salt Lake City, UT 84112, USA}
\author{M.~Kertzman}
\affiliation{Department of Physics and Astronomy, DePauw University, Greencastle, IN 46135-0037, USA}
\author{D.~Kieda}
\affiliation{Department of Physics and Astronomy, University of Utah, Salt Lake City, UT 84112, USA}
\author{M.~Krause}
\affiliation{DESY, Platanenallee 6, 15738 Zeuthen, Germany}
\author{F.~Krennrich}
\affiliation{Department of Physics and Astronomy, Iowa State University, Ames, IA 50011, USA}
\author{S.~Kumar}
\affiliation{Department of Physics and Astronomy and the Bartol Research Institute, University of Delaware, Newark, DE 19716, USA}
\author{M.~J.~Lang}
\affiliation{School of Physics, National University of Ireland Galway, University Road, Galway, Ireland}
\author{T.~T.Y.~Lin}
\affiliation{Physics Department, McGill University, Montreal, QC H3A 2T8, Canada}
\author{S.~McArthur}
\affiliation{Department of Physics and Astronomy, Purdue University, West Lafayette, IN 47907, USA}
\author{P.~Moriarty}
\affiliation{School of Physics, National University of Ireland Galway, University Road, Galway, Ireland}
\author{R.~Mukherjee}
\affiliation{Department of Physics and Astronomy, Barnard College, Columbia University, NY 10027, USA}
\author{S.~O'Brien}
\affiliation{School of Physics, University College Dublin, Belfield, Dublin 4, Ireland}
\author{R.~A.~Ong}
\affiliation{Department of Physics and Astronomy, University of California, Los Angeles, CA 90095, USA}
\author{A.~N.~Otte}
\affiliation{School of Physics and Center for Relativistic Astrophysics, Georgia Institute of Technology, 837 State Street NW, Atlanta, GA 30332-0430, USA}
\author{N.~Park}
\email{npark7@wisc.edu}
\affiliation{Enrico Fermi Institute, University of Chicago, Chicago, IL 60637, USA}
\affil{The Wisconsin IceCube Particle Astrophysics Center and Department of Physics, University of Wisconsin-Madison, Madison, WI, USA}
\author{A.~Petrashyk}
\affiliation{Physics Department, Columbia University, New York, NY 10027, USA}
\author{M.~Pohl}
\affiliation{Institute of Physics and Astronomy, University of Potsdam, 14476 Potsdam-Golm, Germany}
\affiliation{DESY, Platanenallee 6, 15738 Zeuthen, Germany}
\author{E.~Pueschel}
\affiliation{DESY, Platanenallee 6, 15738 Zeuthen, Germany}
\author{J.~Quinn}
\affiliation{School of Physics, University College Dublin, Belfield, Dublin 4, Ireland}
\author{K.~Ragan}
\affiliation{Physics Department, McGill University, Montreal, QC H3A 2T8, Canada}
\author{P.~T.~Reynolds}
\affiliation{Department of Physical Sciences, Cork Institute of Technology, Bishopstown, Cork, Ireland}
\author{G.~T.~Richards}
\affiliation{School of Physics and Center for Relativistic Astrophysics, Georgia Institute of Technology, 837 State Street NW, Atlanta, GA 30332-0430, USA}
\author{E.~Roache}
\affiliation{Fred Lawrence Whipple Observatory, Harvard-Smithsonian Center for Astrophysics, Amado, AZ 85645, USA}
\author{C.~Rulten}
\affiliation{School of Physics and Astronomy, University of Minnesota, Minneapolis, MN 55455, USA}
\author{I.~Sadeh}
\affiliation{DESY, Platanenallee 6, 15738 Zeuthen, Germany}
\author{M.~Santander}
\affiliation{Department of Physics and Astronomy, University of Alabama, Tuscaloosa, AL 35487, USA}
\author{G.~H.~Sembroski}
\affiliation{Department of Physics and Astronomy, Purdue University, West Lafayette, IN 47907, USA}
\author{K.~Shahinyan}
\affiliation{School of Physics and Astronomy, University of Minnesota, Minneapolis, MN 55455, USA}
\author{I.~Sushch}
\affiliation{DESY, Platanenallee 6, 15738 Zeuthen, Germany}
\author{J.~Tyler}
\affiliation{Physics Department, McGill University, Montreal, QC H3A 2T8, Canada}
\author{S.~P.~Wakely}
\affiliation{Enrico Fermi Institute, University of Chicago, Chicago, IL 60637, USA}
\author{A.~Weinstein}
\affiliation{Department of Physics and Astronomy, Iowa State University, Ames, IA 50011, USA}
\author{R.~M.~Wells}
\affiliation{Department of Physics and Astronomy, Iowa State University, Ames, IA 50011, USA}
\author{P.~Wilcox}
\affiliation{Department of Physics and Astronomy, University of Iowa, Van Allen Hall, Iowa City, IA 52242, USA}
\author{A.~Wilhelm}
\affiliation{Institute of Physics and Astronomy, University of Potsdam, 14476 Potsdam-Golm, Germany}
\affiliation{DESY, Platanenallee 6, 15738 Zeuthen, Germany}
\author{D.~A.~Williams}
\affiliation{Santa Cruz Institute for Particle Physics and Department of Physics, University of California, Santa Cruz, CA 95064, USA}
\author{T.~J~Williamson}
\affiliation{Department of Physics and Astronomy and the Bartol Research Institute, University of Delaware, Newark, DE 19716, USA}
\author{B.~Zitzer}
\affiliation{Physics Department, McGill University, Montreal, QC H3A 2T8, Canada}
\collaboration{(The VERITAS Collaboration)}

\author{S.~Abdollahi}
\affiliation{Department of Physical Sciences, Hiroshima University, Higashi-Hiroshima, Hiroshima 739-8526, Japan}
\author{M.~Ajello}
\affiliation{Department of Physics and Astronomy, Clemson University, Kinard Lab of Physics, Clemson, SC 29634-0978, USA}
\author{L.~Baldini}
\affiliation{Universit\`a di Pisa and Istituto Nazionale di Fisica Nucleare, Sezione di Pisa I-56127 Pisa, Italy}
\author{G.~Barbiellini}
\affiliation{Istituto Nazionale di Fisica Nucleare, Sezione di Trieste, I-34127 Trieste, Italy}
\affiliation{Dipartimento di Fisica, Universit\`a di Trieste, I-34127 Trieste, Italy}
\author{D.~Bastieri}
\affiliation{Istituto Nazionale di Fisica Nucleare, Sezione di Padova, I-35131 Padova, Italy}
\affiliation{Dipartimento di Fisica e Astronomia ``G. Galilei'', Universit\`a di Padova, I-35131 Padova, Italy}
\author{R.~Bellazzini}
\affiliation{Istituto Nazionale di Fisica Nucleare, Sezione di Pisa, I-56127 Pisa, Italy}
\author{B.~Berenji}
\affiliation{California State University, Los Angeles, Department of Physics and Astronomy, Los Angeles, CA 90032, USA}
\author{E.~Bissaldi}
\affiliation{Dipartimento di Fisica ``M. Merlin" dell'Universit\`a e del Politecnico di Bari, I-70126 Bari, Italy}
\affiliation{Istituto Nazionale di Fisica Nucleare, Sezione di Bari, I-70126 Bari, Italy}
\author{R.~D.~Blandford}
\affiliation{W. W. Hansen Experimental Physics Laboratory, Kavli Institute for Particle Astrophysics and Cosmology, Department of Physics and SLAC National Accelerator Laboratory, Stanford University, Stanford, CA 94305, USA}
\author{R.~Bonino}
\affiliation{Istituto Nazionale di Fisica Nucleare, Sezione di Torino, I-10125 Torino, Italy}
\affiliation{Dipartimento di Fisica, Universit\`a degli Studi di Torino, I-10125 Torino, Italy}
% ...
\collaboration{(The \textit{Fermi}-LAT Collaboration)}

\author{R.~Alfaro}
\affiliation{Instituto de F\'{\i}sica, Universidad Nacional Aut\'onoma de M{\'e}xico, Mexico City, Mexico}
\author{C.~\'Alvarez}
\affiliation{Universidad Aut\'onoma de Chiapas, Tuxtla Guti{\'e}rrez, Chiapas, Mexico}
\author{J.D.~\'Alvarez}
\affiliation{Universidad Michoacana de San Nicol\'as de Hidalgo, Morelia, Mexico}
\author{R.~Arceo}
\affiliation{Universidad Aut\'onoma de Chiapas, Tuxtla Guti{\'e}rrez, Chiapas, Mexico}
\author{J.C.~Arteaga-Vel\'azquez}
\affiliation{Universidad Michoacana de San Nicol\'as de Hidalgo, Morelia, Mexico}
\author{D.~Avila Rojas}
\affiliation{Instituto de F\'{\i}sica, Universidad Nacional Aut\'onoma de M{\'e}xico, Mexico City, Mexico}
\author{H.A.~Ayala Solares}
\affiliation{Department of Physics, Pennsylvania State University, University Park, PA, USA}
\author{A.~Becerril}
\affiliation{Instituto de F\'{\i}sica, Universidad Nacional Aut\'onoma de M{\'e}xico, Mexico City, Mexico}
\author{E.~Belmont-Moreno}
\affiliation{Instituto de F\'{\i}sica, Universidad Nacional Aut\'onoma de M{\'e}xico, Mexico City, Mexico}
\author{S.Y.~BenZvi}
\affiliation{Department of Physics \& Astronomy, University of Rochester, Rochester, NY, USA}
% ...
\collaboration{(The HAWC Collaboration)}

\begin{abstract}
The HAWC (High Altitude Water Cherenkov) collaboration recently published their 2HWC catalog, listing 39 very high energy (VHE; $>$100~GeV) gamma-ray sources based on 507 days of observation. Among these, there are nineteen sources that are not associated with previously known TeV sources. We have studied fourteen of these sources without known counterparts with VERITAS and \textit{Fermi}-LAT. VERITAS detected weak gamma-ray emission in the 1~TeV--30~TeV band in the region of DA\,495, a pulsar wind nebula coinciding with 2HWC\,J1953+294, confirming the discovery of the source by HAWC. We did not find any counterpart for the selected fourteen new HAWC sources from our analysis of \textit{Fermi}-LAT data for energies higher than 10 GeV. During the search, we detected GeV gamma-ray emission coincident with a known TeV pulsar wind nebula, SNR\,G54.1+0.3 (VER\,J1930+188), and a 2HWC source, 2HWC\,J1930+188. The fluxes for isolated, steady sources in the 2HWC catalog are generally in good agreement with those measured by imaging atmospheric Cherenkov telescopes. However, the VERITAS fluxes for SNR\,G54.1+0.3, DA\,495, and TeV\,J2032+4130 are lower than those measured by HAWC and several new HAWC sources are not detected by VERITAS. This is likely due to a change in spectral shape, source extension, or the influence of diffuse emission in the source region. 
\end{abstract}

\section{Introduction}
Gamma-ray astronomy can be performed using a variety of techniques, each with different strengths and weaknesses. Direct detection of gamma rays is possible with  satellite-based instrumentation, such as the Large Area Telescope (LAT) on board the \textit{Fermi Gamma-Ray Space Telescope}~\citep{2009ApJ...697.1071A}. This provides low background observations over a wide field of view, covering about 20$\%$ of the sky at any given time and scanning the whole sky every three hours. However, due to the physical size limitations imposed upon satellite-based instruments, the effective area is generally smaller than 1 $\textnormal{m}^{2}$, leading to a sensitivity that peaks at a few GeV. Above 100~GeV, ground-based observatories are best suited to study the emission, thanks to their large effective collection area when compared to space experiments. Ground-based imaging atmospheric Cherenkov telescope (IACT) arrays, such as VERITAS~\citep{2002APh....17..221W}, observe the Cherenkov light generated by particle showers in the atmosphere, while air shower arrays, such as the High Altitude Water Cherenkov (HAWC) observatory~\citep{2013APh....50...26A}, sample the air shower particles at ground level. 
IACTs offer the best instantaneous sensitivity thanks to their large effective collection area ($\sim$$\textnormal{10}^5 \UU{m}{2}$) and excellent rejection of the cosmic-ray background. However, observations require clear, dark skies, limiting the duty cycle to $\lesssim$20$\%$, and gamma-ray sources must be contained within the field of view of the telescope, which is at present $\lesssim$5$\degree$ diameter. Air shower arrays for gamma-ray observations provide lower instantaneous sensitivity than IACTs, but they can operate continuously with an instantaneous field of view of the telescope covering $\sim$15\% of the sky. Sensitive, unbiased surveys for a large portion of the sky can be conducted over the lifetime of air shower arrays. 

The angular and energy resolution of each of the three techniques, which allow one to study and to understand astrophysical gamma-ray sources in detail, are complementary. For example, the good angular resolution of IACTs allows us to resolve the detailed morphology of spatially extended sources and to identify the counterparts of sources in complex regions. The limited field of view, however, restricts the size of the emission region that can be studied. Compared to this, satellite-based instruments and air shower arrays can provide good measurements of highly extended sources. Satellite-based instruments provide energy resolution better than 15$\%$ for gamma rays with energies above several hundreds of MeV up to around 1~TeV. Above 1~TeV, IACTs provide the best energy resolution (generally about 20$\%$). Combined with their large effective areas and sensitivities, IACTs thus can measure detailed features of the spectral energy distribution (SED) of sources. Air shower arrays' energy resolution is worse than that of IACTs. The large and relatively uniform exposure time of air shower array measurements, however, can provide good high-energy measurements above tens of TeV for a large portion of the sky, increasing the dynamic range of the measurements and allowing the study of spectral changes at the highest energies. The most powerful approach, therefore, is to combine observations from all three methods. Only a few examples of this exist~\citep{2014ApJ...787..166A,2014ApJ...788...78A} due to limited overlapping source catalogs.

In this paper, we describe the results of observations of newly discovered HAWC sources with the VERITAS IACT array and the LAT on board the \textit{Fermi Gamma-Ray Space Telescope}.
Fully completed in March 2015, HAWC has recently released a catalog, 2HWC~\citep{2017ApJ...843...40A}. Compared to the previous very high energy (VHE; $>$100~GeV) surveys performed by Milagro~\citep{2007ApJ...664L..91A} and ARGO-YBJ~\citep{2013ApJ...779...27B}, HAWC provides more than an order of magnitude better sensitivity~\citep{2017ApJ...843...39A}. The 2HWC catalog contains 39 sources, twenty of which are associated with known astrophysical objects including active galactic nuclei, pulsar wind nebulae (PWNs), and supernova remnants (SNRs). The remaining nineteen sources in the catalog have not previously been identified as TeV gamma-ray emitters, providing promising new targets for follow-up observations with IACTs and space-based gamma-ray observatories.

\section{Target selection and observations}
\subsection{HAWC and the 2HWC catalog}
HAWC is an air shower array in operation in central Mexico, consisting of 300 water-filled, light-tight tanks, each instrumented with 4 photomultiplier tubes (PMTs). The PMTs in each tank detect the Cherenkov light emitted by particles from the air showers. Relative timing between PMTs allows the reconstruction of the direction of the shower plane, and hence of the primary particle. HAWC operates 24 hours per day with an instantaneous field of view of $\sim$2~sr. The Earth's rotation enables HAWC to observe 2/3 of the sky every day. HAWC was inaugurated on 2015 March 20, but its modular design allowed partial operation before then. HAWC is sensitive to gamma rays from 100~GeV to 100~TeV, with a one-year survey sensitivity to detect sources with an average flux corresponding to 5--10$\%$ of the flux of the Crab Nebula across most of the Northern sky. The data presented here were collected between 2014 November 26 and 2016 June 2, amounting to a livetime of 507 days. 

Details of the HAWC analysis techniques, including a study of systematic uncertainties, are described in~\citet{2017ApJ...843...39A}. Data analysis is performed in bins, $\mathcal{B}_{i}$,$i=1\ldots9$, which correspond to the fraction of PMTs, $f_{\textnormal{hit}}$, reporting a signal for a given event. The energy of the primary gamma ray is correlated with $f_{\textnormal{hit}}$, so bin $\mathcal{B}_{i}$ is used as an energy proxy. Gamma rays are discriminated from cosmic ray background using the variance of the charge distribution detected in each air shower, with cuts optimized for each bin $\mathcal{B}$. The angular resolution, defined as a radius that contains 68$\%$ of flux from a point source, depends strongly on the analysis bin, from 1.0$\degree$ for $\mathcal{B}_{1}$ to 0.18$\degree$ for $\mathcal{B}_{9}$. A maximum likelihood method was used to reconstruct the spectrum. It takes into account the point spread function (PSF) for each $\mathcal{B}_{i}$ and compares the expected number of events, given the experimentally measured background, to a given spectral hypothesis.

In the 507-day operation dataset, sources were identified in four all-sky maps. One map was optimized for point sources, while the remaining three maps were optimized for extended sources of diameters 0.5$\degree$, 1$\degree$, and 2$\degree$. A \emph{top hat} distribution with the given diameter was used to smooth the skymap for each extended source search. For each map, a test statistic (TS) value was calculated based on the ratio of the likelihood with a single source model and with a pure background model. The 2HWC analysis required the TS value of a source to be higher than 25. The expected number of false detections for this analysis is 0.5. The analysis for the 2HWC catalog was carried out for all of the sources based on a hypothesis of a simple power law spectral distribution, $dN/dE=\textnormal{N}_{0}\textnormal{E}^{-\gamma}$, where $\textnormal{N}_{0}$ is a normalization factor and $\gamma$ is a spectral index. Spectral fits were performed assuming a point-source morphology for all sources identified in the point source all-sky map. For sources identified in the extended maps, a \emph{top hat} morphology was assumed with a size matching with the map in which the extended source or candidate was found to be most significant. Of the 39 HAWC sources, nineteen are located more than 0.5$\degree$ away from previously known TeV sources, presenting a group of newly detected TeV sources. These sources generally have low TS values compared to other 2HWC sources that are associated with known TeV sources. The average value of the spectral indices is 2.6 with the spectral indices ranging from 1.5 to 3.3. We use \textit{Fermi}-LAT and VERITAS data to look for the counterparts of these HAWC sources that do not have clear associations with previous detected TeV sources\footnote{http://tevcat.uchicago.edu}. 

\subsection{Observations}\label{Sec:2HWC_Observation}
We searched VERITAS archival data collected from 2007 to 2015 for the nineteen 2HWC sources without known counterparts and selected data taken with the pointing of the telescopes offset by less than 1.5$\degree$ from the locations of the HAWC sources. Of these nineteen HAWC sources, eleven locations had been observed by VERITAS prior to 2015 with a total exposure time of 134 hours. In addition to the archival data, VERITAS observed a subset of the HAWC sources during the 2015--2016 and the 2016--2017 seasons. Combining the archival and new data sets, VERITAS observed a total of fourteen out of nineteen new sources reported by HAWC. After data quality selections, a total of 218 hours of data was analyzed for the study. Detailed information about the sources is shown in Table~\ref{Tbl:ListOfSources}. For each source listed in this table, the 2HWC identifier is provided, along with the map in which it was identified (PS for point source, 0.5$\degree$ or 1$\degree$ for extended source) and J2000 Right Ascension (RA) and declination (Dec). The 1$\sigma$ statistical uncertainty of the source position is shown as `Unc'. Sources marked with an asterisk($\ast$) were identified as local maxima in the TS maps but are not as clearly separated from neighboring sources. The table also shows the power law index and differential flux at 7~TeV, $F_7$, reported by HAWC. The flux is reported at 7~TeV because this energy results in the least correlation between spectral index and flux. If the source was identified in an extended source map, then the fit result with an integration radius used for the extended source search is shown. In addition to the catalog values, we present the minimum energy and maximum energy of the central interval that contributes 75\% of the TS for a given source as $E_\mathrm{12.5}$ and $E_\mathrm{87.5}$. The exact values of $E_\mathrm{12.5}$ and $E_\mathrm{87.5}$ depend on both source declination and spectral index, and are determined individually for each source. If $E_\mathrm{87.5}$ exceeds 100~TeV, we only report a lower bound, as the data analysis techniques used for the 2HWC catalog are unable to measure energies higher than approximately this value. The exposure time of VERITAS for each source varies from 1.3 hours to 72 hours as shown in Table~\ref{Tbl:NonDetections}. We analyzed 8.5 years of \textit{Fermi}-LAT data from 2008 August to 2017 February for the study.  

\begin{deluxetable*}{c  c  c  cc c  c c c c}
\tablecaption{List of HAWC sources for the \textit{Fermi}-VERITAS follow-up study. Data for individual sources is taken from the 2HWC catalog~\citep{2017ApJ...843...40A}, with the exception of $E_\mathrm{12.5}$ and $E_\mathrm{87.5}$. Detailed description is in Section \ref{Sec:2HWC_Observation}.\label{Tbl:ListOfSources}}
\tablehead{
\colhead{Name} & \colhead{Id} & \colhead{TS} & \colhead{RA} & \colhead{Dec} & \colhead{Unc.} &\colhead{Index} & \colhead{$F_7$} &  \colhead{$E_{12.5}$} &  \colhead{$E_{87.5}$}\\
\colhead{} & \colhead{} & \colhead{} & \colhead{[$\degree$] } & \colhead{[$\degree$] } & \colhead{[$\degree$] } & \colhead{} & \colhead{[$10^{-15}$ $\textnormal{cm}^{-2}$s$^{-1}$TeV$^{-1}$]} & \colhead{[TeV]} & \colhead{[TeV]} 
}
\startdata
2HWC J0700+143 & 1 & 29 & 105.12 & 14.32 & 0.80 &  2.17$\pm$0.16 & 13.8 $\pm$ 4.2 & 2.3 & $>$100 \\
2HWC J0819+157 & 0.5 & 30.7 & 124.98 & 15.79 & 0.17 & 1.50$\pm$0.67 & 1.6 $\pm$ 3.1 & 25 & $>$100 \\
2HWC J1040+308 & 0.5 & 26.3 & 160.22& 30.87 & 0.22 & 2.08$\pm$0.25 & 6.6 $\pm$ 3.5 & 6 & $>$ 100 \\
2HWC J1852+013$^{\ast}$ & PS & 71.4 & 283.01 & 1.38& 0.13 & 2.90$\pm$0.10 & 18.2 $\pm$ 2.3 & 0.4 & 50\\
2HWC J1902+048$^{\ast}$ & PS & 31.7 &  285.51& 4.86& 0.18 & 3.22$\pm$0.16 & 8.3 $\pm$ 2.4 & 0.3 & 11\\
2HWC J1907+084$^{\ast}$ & PS & 33.1 &  286.79& 8.50& 0.27 & 3.25$\pm$0.18 & 7.3 $\pm$ 2.5 & 0.18 & 10\\
2HWC J1914+117$^{\ast}$ & PS & 33 &  288.68& 11.72& 0.13 & 2.83$\pm$0.15 & 8.5 $\pm$ 1.6 & 0.5 & 42\\
2HWC J1928+177 & PS & 65.7 & 292.15 &17.78& 0.07 & 2.56$\pm$0.14 & 10.0 $\pm$ 1.7 & 0.9 & 86\\
2HWC J1938+238 & PS & 30.5 & 294.74 &23.81& 0.13 & 2.96$\pm$0.15 & 7.4 $\pm$ 1.6 & 0.3  & 29\\
2HWC J1949+244 & 1& 34.9 & 297.42 & 24.46& 0.71 & 2.38$\pm$0.16 & 19.4 $\pm$ 4.2 & 1.1 & $>$100\\
2HWC J1953+294 & PS & 30.1 & 298.26 & 29.48& 0.24 & 2.78$\pm$0.15 & 8.3 $\pm$ 1.6 & 0.6 & 55\\ 
2HWC J1955+285 & PS & 25.4 & 298.83 &28.59& 0.14 & 2.40$\pm$0.24 & 5.7 $\pm$ 2.1 & 1.6 & $>$100\\
2HWC J2006+341 & PS & 36.9 & 301.55 &34.18& 0.13 & 2.64$\pm$0.14 & 9.6 $\pm$ 1.9 & 1.0 & 86\\
2HWC J2024+417$^{\ast}$ & PS & 28.4 & 306.04 & 41.76 & 0.20 & 2.74$\pm$0.17 & 12.4 $\pm$ 2.6 & 1.0 & 100\\
\enddata
\end{deluxetable*}

\section{Follow-up instruments and analyses}\label{Sec:Analysis}
\subsection{VERITAS}
VERITAS is an array of four IACTs located at the Fred Lawrence Whipple Observatory in southern Arizona~\citep{2002APh....17..221W}. 
Each telescope has a tessellated 12-meter diameter reflector which is used to collect the Cherenkov light generated by gamma-ray-initiated particle cascades (or air showers) in the Earth's atmosphere. A camera composed of 499 PMTs is installed at the focal plane of each reflector and used to record an image of the cascades. VERITAS is designed to detect gamma rays from an energy of 85~GeV to energies higher than 30~TeV, over a field of view with a diameter of 3.5$\degree$. Since the beginning of full array operations in 2007, the sensitivity of VERITAS has been improved by two major upgrades~\citep{2015arXiv150807070P}---the relocation of one telescope in 2009~\citep{2009arXiv0912.3841P} and the upgrade of the camera with high quantum efficiency PMTs in 2012~\citep{2013arXiv1308.4849D}. With its current configuration, VERITAS can detect a point source with 1$\%$ of the flux from the Crab Nebula within 25 hours, and has an angular resolution better than 0.1$\degree$ at 1~TeV. 

The VERITAS analysis begins with standard calibration and image cleaning procedures, after which each image is parameterized geometrically. A standard Hillas moment parameterization is used for this study~\citep{1985ICRC....3..445H}; a detailed description of the VERITAS data analysis procedure can be found in \cite{2008ICRC....3.1325D}. Selection cuts are then applied to the data in order to discriminate gamma-ray-initiated events from the otherwise overwhelming background of cosmic-ray-initiated cascades. The choice of the optimum gamma-ray selection cuts depends upon the assumed properties of the source candidate, provided by HAWC in this case. The peak sensitivity of HAWC is located in the multi-TeV energy range. For source regions on which the VERITAS exposure is larger than $10\U{hours}$, we therefore apply strict gamma-ray selection cuts that are optimized for objects with a hard spectral index, and with a weak signal ($\sim$1\% of the steady Crab Nebula flux). These cuts provide the best sensitivity for gamma rays with energies higher than $\sim$1~TeV~\citep{2015arXiv150807070P} while still providing good sensitivity down to 300--600~GeV. For exposures shorter than $10\U{hours}$, we chose a less strict set of cuts suitable for stronger sources ($3\%\U{Crab}$), which retain more gamma rays at the expense of higher background. This is justified by the fact that a weaker source would be below the sensitivity of VERITAS in such a short exposure. Additional gamma-ray discrimination is achieved by cutting on $\theta$, the angular distance between a test position on the sky and the shower arrival direction. The angular extension of the 2HWC sources is not well defined.
Also, a source which appears point-like to HAWC may be an extended source for VERITAS because of its smaller PSF. We therefore applied two sets of angular cuts to the VERITAS search: one for a point-like source ($\theta$$<$0.1$\degree$) and the other for a source with moderately large angular extent ($\theta$$<$0.3$\degree$ for 2HWC\,J1953+294 and $\theta$$<$0.23$\degree$ for the rest of the sources). The results described here have been confirmed by two independent analysis chains~\citep{2008ICRC....3.1385C,2017arXiv170804048M}.

\subsection{\textit{Fermi}-LAT}

The LAT is a high-energy gamma-ray telescope which detects photons from 20~MeV to higher than 500~GeV~\citep{2009ApJ...697.1071A}. Since the launch of the spacecraft in 2008 June, the event-level analysis has been improved based on our increased knowledge of the detectors. The latest version, dubbed Pass 8~\citep{2013arXiv1303.3514A}, offers a greater acceptance compared to previous LAT data and an improved PSF (with a 68\% containment radius less than 0.2$\degree$ above 10~GeV that is nearly constant with increasing energy). Together, these two factors significantly improve the detection of sources at energies above 10~GeV. To search for a counterpart of HAWC sources with \textit{Fermi}-LAT data, we decided to focus on these high-energy events. By limiting the analysis to high energies we reduce contamination from diffuse Galactic emission. All but three of the HAWC sources investigated here lie within 2.5\degree of the Galactic plane (the sources outside the plane lie at a right ascension of less than 11 hours). The LAT energy selection also limits confusion from gamma-ray pulsars in the Galactic plane because gamma-ray pulsar spectra typically roll over in the few GeV energy range. For energies lower than 10 GeV, we utilized publicly available information from the \textit{Fermi}-LAT Third Source Catalog (3FGL)~\citep{2015ApJS..218...23A}.  

As the starting point for our analysis we used a model based on the Third Catalog of Hard \textit{Fermi}-LAT Sources~\citep[3FHL;][]{2017ApJS..232...18A} and the \textit{Fermi} Galactic Extended Source (FGES) catalog~\citep{2017arXiv170200476T}. The 3FHL catalog contains sources detected between 10~GeV and 2~TeV using 7 years of Pass 8 data, while the FGES catalog focuses on the study of extended sources discovered in the same energy range using 6 years of Pass 8 data. For the unassociated HAWC sources discussed here, there are no spatially coincident LAT sources detected in either catalog. To search for new LAT counterparts and to place upper limits at the position of these new HAWC sources, we analyzed 8.5 years of LAT data from 2008 August to 2017 February using Pass 8 SOURCE photons with reconstructed energies in the 10~GeV to 0.9~TeV range. To limit contamination from Earth limb photons we excluded photons at zenith angles larger than 105$\degree$. The \textit{Fermi} Science Tools v10r01p01, the instrument response functions (IRFs) P8R2 SOURCE V6 and the fermipy v0.13 analysis package~\citep{2017arXiv170709551W} were used for this analysis.  To model the diffuse backgrounds we used the standard templates for isotropic and Galactic interstellar emission \footnote{Galactic IEM: gll\_iem\_v06.fits, Isotropic: iso\_P8R2\_SOURCE\_V6\_v06.txt. Please see: \url{http://fermi.gsfc.nasa.gov/ssc/data/access/lat/BackgroundModels.html} } developed by the LAT collaboration.

The analysis has proceeded as follows. A 10$\degree$ square region centered around each HAWC source is fit using a binned likelihood formalism. We first place a point source at the location of the HAWC source. The source is modeled by a power law energy spectrum with freely fit index and normalization throughout the fit. Using fermipy, we attempt to localize any putative gamma-ray source around the HAWC position. A source is considered detected if it has a TS greater than 25, defined as TS $= -2 \ln(L_{\textnormal{max,0}}/L_{\textnormal{max,1}})$, where $L_{\textnormal{max,0}}$ is the maximum likelihood value for a model without an additional source (the ``null hypothesis'') and $L_{\textnormal{max,1}}$ is the maximum likelihood value for a model with the additional source at a specified location. We repeat this procedure with a uniformly illuminated disk with an initial radius of 0.2$\degree$. The localization and extension of the disk are both fit to search for a possible spatially extended LAT counterpart. 

In the event that no significant point-like or extended source is detected, integral upper limits are computed at the 99\% confidence level using a semi-Bayesian method provided in the \textit{Fermi} Science Tools. We compute these upper limits for three different power law spectral indices: 2.0, 3.0, and the index reported in 2HWC. A pivot energy of 20 GeV is assumed for the conversion to a differential flux limit as the LAT has a decreasing sensitivity with increasing energy above 10 GeV.
We compute upper limits assuming either a point source model or an extended source. For the extended source model, we choose an extension equal to that estimated in 2HWC or the approximate localization of a point source by HAWC ($\sim$0.2$\degree$).

\section{Results} 
\begin{figure*}[ht!]
  \centering
  \includegraphics[scale=0.8]{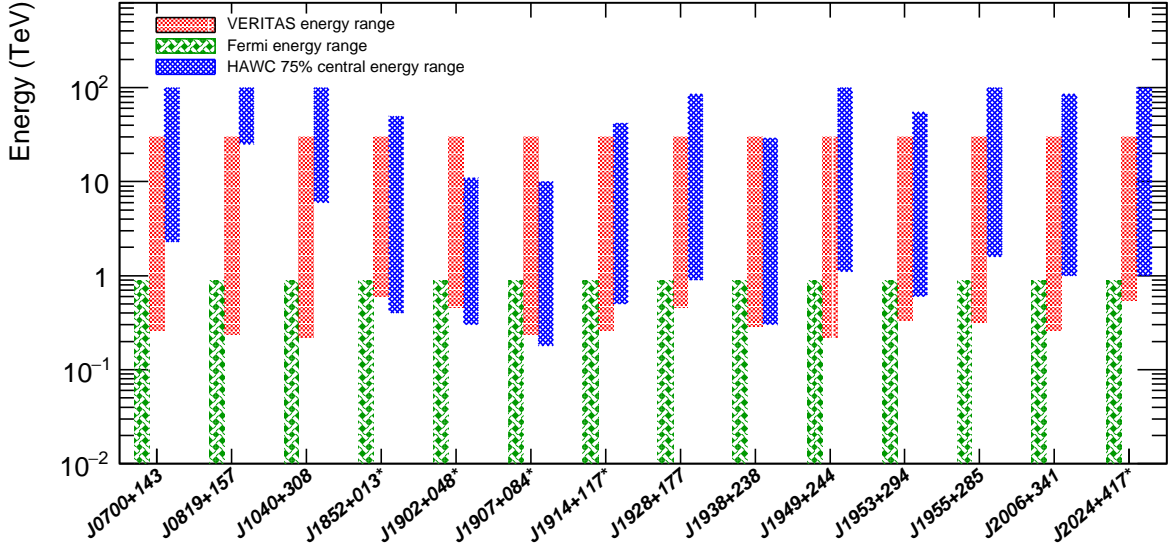}
\caption{Energy range comparisons between the three instruments. The energy coverage for each instrument is shown as the green filled block for \textit{Fermi}-LAT, red filled block for VERITAS, and blue filled block for HAWC.} 
\label{Fig:EnergyRangeComparisons}
\end{figure*}

\begin{figure*}[ht!]
  \centering
  \includegraphics[scale=0.8]{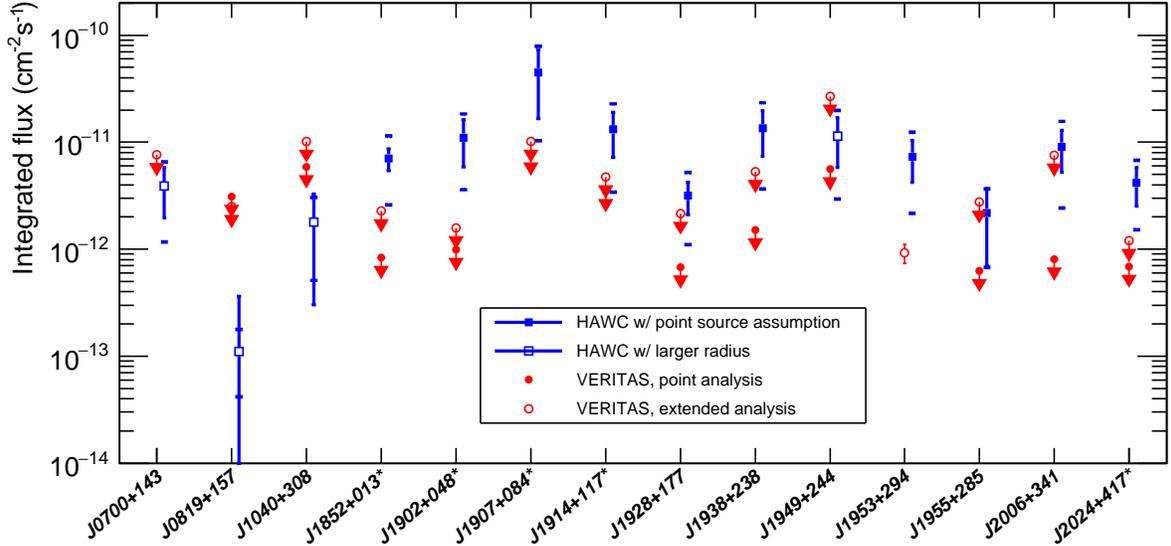}
\caption{Integrated photon flux comparison between HAWC and VERITAS. For the comparison, the energy range of the VERITAS analysis shown with red filled blocks in Figure~\ref{Fig:EnergyRangeComparisons} was used. The error bars for the HAWC flux are statistical errors derived from the propagation of statistical errors on the normalization factors and indices. The systematic error of the HAWC flux is shown with blue horizontal ticks. VERITAS flux upper limits are calculated assuming a power law distribution with HAWC's spectral index for each source. }  
\label{Fig:FluxComparisons}
\end{figure*}

The energy ranges of \textit{Fermi}-LAT, VERITAS, and HAWC for this study are shown in Figure~\ref{Fig:EnergyRangeComparisons}. The \textit{Fermi}-LAT energy ranges are determined by event selections, and they are the same for all of the selected sources. The VERITAS energy range varies from source to source due to different gamma-ray selection cuts used for the analyses and different observational conditions. The energy threshold value of the analysis sets the minimum energy range, while the maximum energy range is set to 30~TeV to choose events reconstructed with energy bias smaller than 10$\%$. For HAWC, the energy ranges correspond to the central 75$\%$ of energies contributing to the TS for the source. 

\textit{Fermi}-LAT did not detect counterparts for any of the fourteen HAWC sources considered in this study, in either the point source or extended source searches. Only one gamma-ray source was detected by VERITAS out of fourteen selected sources. The detected gamma-ray source was found in the region of 2HWC\,J1953+294.

\subsection{Sources not detected by \textit{Fermi}-LAT and VERITAS}\label{Sec:Sources-not-VERITAS}
For sources that are not detected by \textit{Fermi}-LAT and VERITAS, upper limits are calculated at the 99\% confidence level by using three spectral indices : 2.0, 3.0, and the spectral index reported by HAWC. Spectral indices of 2.0 and 3.0 are selected to consider possible spectral changes from HAWC's energy range to the lower energy band.
Upper limits from \textit{Fermi}-LAT and VERITAS for the non-detected thirteen sources are shown in Table~\ref{Tbl:NonDetections}. The upper limits for \textit{Fermi}-LAT and VERITAS are calculated at the location of each HAWC source over energy ranges shown in Figure~\ref{Fig:EnergyRangeComparisons}. 

To compare the VERITAS upper limits with the flux measurement of HAWC, we calculate the integrated flux from each source in the VERITAS energy range, using the spectral information measured by HAWC. The result is shown in Figure~\ref{Fig:FluxComparisons}. Error bars for the HAWC flux estimates were derived with error propagation using the statistical errors of the flux normalization factors at 7~TeV and the indices. The systematic errors of the HAWC flux shown as brackets were calculated with a flux normalization error of 50\% and an index error of 0.2~\citep{2017ApJ...843...39A}. 

For the point source search, the VERITAS upper limits are lower than the expected flux estimated by using HAWC's spectral information by more than 1$\sigma$ for most of the sources except three---2HWC\,J0819+157, 2HWC\,J1040+308, and 2HWC\,J1949+244. For the extended source search, there are three sources, 2HWC\,J1852+013$^{\ast}$, 2HWC\,J1902+048$^{\ast}$, and 2HWC\,J2024+417$^{\ast}$, for which the VERITAS upper limits using the integration radius of 0.23$\degree$ are lower than the flux estimated by HAWC by more than 1$\sigma$. The upper limits of the other ten sources are consistent with the HAWC flux estimation within 1$\sigma$. VERITAS detected gamma-ray emission from 2HWC\,J1953+294, but there is a discrepancy in the flux estimation between VERITAS and HAWC (discussed in section~\ref{Sec:DA495region}.) 

We also compared the upper limits of \textit{Fermi}-LAT with the extrapolation of HAWC's spectra to \textit{Fermi}-LAT energy ranges. Because the \textit{Fermi}-LAT energy range is much lower than HAWC's energy range, the extrapolation has larger uncertainties. In this study, we found that both point and extended source upper limits calculated with \textit{Fermi}-LAT are lower than HAWC's flux extrapolation by more than 1$\sigma$ for five sources--2HWC\,J1852+013$^{\ast}$, 2HWC\,J1914+117$^{\ast}$, 2HWC\,J1928+177, 2HWC\,J1938+238, and 2HWC\,J1953+294. For 2HWC\,J2006+341, only the point source upper limit is lower than HAWC's flux estimation. 

Individual SEDs of selected 2HWC sources are shown in the appendix together with the upper limits from \textit{Fermi}-LAT and VERITAS. 

\subsection{SNR\,G54.1+0.3 region}\label{Sec:SNRG54.1}
The first region that we discuss in detail contains two HAWC sources--2HWC\,J1930+188 and 2HWC\,J1928+177. 2HWC\,J1930+188 is coincident with a TeV source previously identified by VERITAS, VER\,J1930+188~\citep{2010ApJ...719L..69A} while 2HWC\,J1928+177 is newly identified by HAWC. The VERITAS excess counts map of this region is shown in Figure~\ref{Fig:G54region_skymap}. 

\tabletypesize{\scriptsize}
\begin{longrotatetable}
\begin{deluxetable*}{l  c  ccc  cccc  c}
\tablecaption{Sources not detected by VERITAS and $\textit{Fermi}$-LAT.\label{Tbl:NonDetections}}
\tablehead{
\colhead{} & \colhead{} &\multicolumn{3}{c}{\textit{Fermi}-LAT} &\multicolumn{4}{c}{VERITAS} &\colhead{HAWC} \\
\colhead{} &\colhead{} &\colhead{Point source} &\colhead{Test} &\colhead{Extended source} &\colhead{Exposure} &\colhead{$E_{thr}$} &\colhead{Point source} &\colhead{Extended source} &\colhead{Integrated} \\
\colhead{Source name} &\colhead{Index} &\colhead{upper limit} &\colhead{radius} &\colhead{upper limit} &\colhead{time} &\colhead{} &\colhead{upper limit} &\colhead{upper limit} &\colhead{flux$\dagger$} \\
  \colhead{} & \colhead{}  & \colhead{[$10^{-11}$ $\textnormal{cm}^{-2}\textnormal{s}^{-1}$]} & \colhead{[$\degree$]} & \colhead{[$10^{-11}$ $\textnormal{cm}^{-2}\textnormal{s}^{-1}$]} & \colhead{[hour]} & \colhead{[GeV]} & \colhead{[$10^{-12}$ $\textnormal{cm}^{-2}\textnormal{s}^{-1}$]} & \colhead{[$10^{-12}$ $\textnormal{cm}^{-2}\textnormal{s}^{-1}$]} &
\colhead{[$10^{-12}$ $\textnormal{cm}^{-2}\textnormal{s}^{-1}$]} 
}
\startdata
	\multirow{3}{*}{2HWC J0700+143} 
	& 2.0 &  4.1 & \multirow{3}{*}{1.0} & 6.1 & \multirow{3}{*}{5.6} & 290 & \ldots & 6.8 & \multirow{3}{*}{3.9 $\pm$ 1.9}\\
	& 2.17 & 4.0 & & 6.2 & & 260 & \ldots & 7.7 & \\
	& 3.0 &  3.4 & & 5.7 & & 240 & \ldots & 9.0 & \\
\multirow{3}{*}{2HWC J0819+157} 
	& 2.0 &  1.8 & \multirow{3}{*}{0.5} & 2.3 & \multirow{3}{*}{4.5} & 220 & 2.4 & 2.7 & \multirow{3}{*}{0.11 $\pm$ 0.25}\\
	& 1.5 & 1.6 & & 1.9 & & 240 & 3.1 & 2.5 &\\
	& 3.0 &  2.0 & & 2.7 & & 200 & 3.0 & 3.3 & \\
	\multirow{3}{*}{2HWC J1040+308} 
	& 2.0 & 1.7 & \multirow{3}{*}{0.5} & 2.0 & \multirow{3}{*}{3.1} & 220 & 5.7 & 10 & \multirow{3}{*}{1.8 $\pm$ 1.5}\\
	& 2.08 & 1.7 & & 2.0 & & \ldots & \ldots & \ldots &\\
	& 3.0 & 1.6 & & 2.0 & & 200 & 7.9 & 13 &\\
	\multirow{3}{*}{2HWC J1852+013$^{\ast}$} 
	& 2.0 & 2.9 & \multirow{3}{*}{0.23} & 3.8 & \multirow{3}{*}{10} & 660 & 0.41 & 1.1 & \multirow{3}{*}{7.1 $\pm$ 1.6}\\
	& 2.9 & 3.0 & & 4.2 & & \ldots & \ldots & \ldots &\\
	& 3.0 & 3.0 & & 4.2 & & 420 & 0.84 & 2.3 &\\
	\multirow{2}{*}{2HWC J1902+048$^{\ast}$} 
	& 2.0 & 3.3 & \multirow{3}{*}{0.23} & 4.57 & \multirow{2}{*}{20} & 550 & 0.79 & 1.6& \multirow{3}{*}{11 $\pm$ 5.1} \\
	& 3.22 & 4.9 & & 6.6 & & 460 & 0.99 & 1.6& \\
	& 3.0 & 4.7 & & 6.3 & & 500 & 0.83 & 1.6 &\\
	\multirow{2}{*}{2HWC J1907+084$^{\ast}$} 
	& 2.0 & 2.3 &  \multirow{3}{*}{0.23} &3.1 & \multirow{2}{*}{4.9} & 290 & 5.1 & 7.1& \multirow{3}{*}{45 $\pm$ 28} \\
	& 3.25 & 2.5 & & 3.5 & & 240 & 7.7 & 10 &\\
	& 3.0 & 2.5 & & 3.5 & & 260 & 6.3 & 10 &\\
    \multirow{3}{*}{2HWC J1914+117$^{\ast}$} 
	& 2.0 & 3.4 & \multirow{3}{*}{0.23} & 3.9 & \multirow{3}{*}{2.2} & 260 & 3.0 & 4.4 & \multirow{3}{*}{13 $\pm$ 6.0} \\
	& 2.83& 3.7 & & 4.5 & & 260 & 3.5 & 4.7 & \\
	& 3.0 & 3.7 & & 4.5 & & 260 & 3.6 & 4.8 & \\
	\multirow{3}{*}{2HWC J1928+177} 
	& 2.0 & 4.9 &  \multirow{3}{*}{0.23} &6.5 & \multirow{3}{*}{44} & 500 & 0.63 & 2.0 & \multirow{3}{*}{3.2 $\pm$ 1.1} \\
	& 2.56 & 5.3 & & 6.9 & & 460 & 0.68 & 2.2 & \\
	& 3.0 & 5.4 & & 7.0 & & 320 & 1.3 & 4.2 & \\
	\multirow{2}{*}{2HWC J1938+238} 
	& 2.0 & 3.4 & \multirow{3}{*}{0.23} & 4.6 & \multirow{2}{*}{2.9} & 320 & 1.3 & 4.5 & \multirow{3}{*}{14 $\pm$ 6.2} \\
	& 2.96& 3.3 & & 4.5 & & 290 & 1.5 & 5.3 & \\
	& 3.0 & 3.3 & & 4.5 & & 260 & 1.8 & 6.4 & \\
     \addlinespace[13pt]
	\multirow{3}{*}{2HWC J1949+244} 
	& 2.0 & 3.5 &  \multirow{3}{*}{1.0} & 12 & \multirow{3}{*}{1.8} & 220 & 5.0 & 24 & \multirow{3}{*}{11 $\pm$ 5.6}\\
	& 2.38 & 3.4 & & 12 & & 220 & 5.6 & 27 &\\
	& 3.0 & 3.4 & & 10 &  & 200 & 7.6 & 36 &\\
	\multirow{3}{*}{2HWC J1955+285} 
	& 2.0 & 2.7 & \multirow{3}{*}{0.23} & 3.4 & \multirow{3}{*}{46} & 320 & 0.61 & 2.7 & \multirow{3}{*}{2.2 $\pm$ 1.5}\\
	& 2.40 & 2.7 & & 3.5 & & 320 & 0.63 & 2.8 & \\
	& 3.0 & 2.7 & & 3.4 & & 320 & 0.62 & 2.7 & \\
	\multirow{3}{*}{2HWC J2006+341} 
	& 2.0 & 3.4 &  \multirow{3}{*}{0.23} & 48 & \multirow{3}{*}{7.0} & 290 & 0.61 & 5.8 & \multirow{3}{*}{9.1 $\pm$ 3.9}\\
	& 2.64& 3.4 & & 47 & & 260 & 0.80 & 7.5 & \\
	& 3.0 & 3.4 & & 46 & & 240 & 0.84 & 7.9 &\\
    	\multirow{3}{*}{2HWC J2024+417$^{\ast}$} 
	& 2.0 & 2.8 &  \multirow{3}{*}{0.23}  & 5.9 & \multirow{3}{*}{40} & 720 & 0.41 & 0.8& \multirow{3}{*}{4.2 $\pm$ 1.7}\\
	& 2.74& 3.2 & & 6.59 & & 550 & 0.69 & 1.2 &\\
	& 3.0 & 3.3 & & 6.54 & & 550 & 0.73 & 1.5 &\\
\enddata
\tablenotetext{\dagger}{The integrated flux shown here for HAWC is calculated by using the spectral shape provided by HAWC's measurement over the energy range provided by VERITAS's analysis for each source.}
\end{deluxetable*}
\end{longrotatetable}

\subsubsection{SNR\,G54.1+0.3}

VERITAS detected a point-like source of TeV gamma-ray emission, VER\,J1930+188, coincident with SNR\,G54.1+0.3 in this region~\citep{2010ApJ...719L..69A}. H.E.S.S. also recently reported a detection of the source, HESS\,J1930+188, in their Galactic plane survey~\citep{2018A&A...612A...1H}. The centroid and integral flux measured by H.E.S.S. agree with VERITAS within statistical and systematic errors. The VERITAS source is associated with SNR\,G54.1+0.3, a PWN surrounding a radio/X-ray pulsar, PSR\,J1930+1852. 

\begin{figure*}[ht!]
  \centering
  \includegraphics[width=0.8\textwidth]{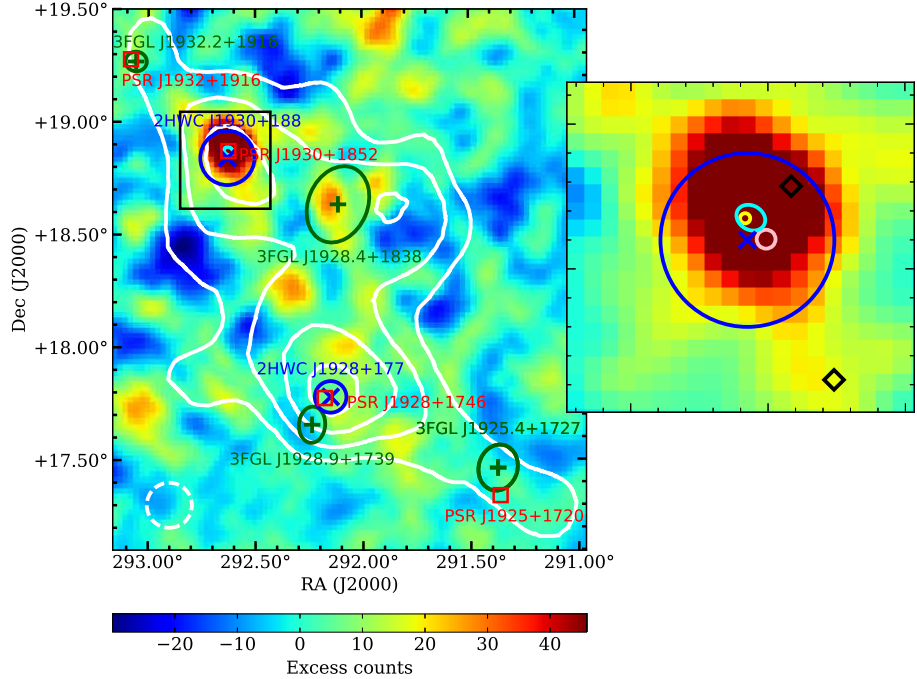}
\caption{VERITAS gamma-ray counts map of the SNR\,G54.1+0.3 region with point source search cuts. The $\theta$ cut used for the study, $\theta<$0.1$^{\circ}$, is shown as a white dashed circle. The extension of the radio emission from SNR\,G54.1+0.3~\citep{2010ApJ...709.1125L} is shown with a cyan ellipse. A zoomed 0.36$^\circ$ by 0.36$^\circ$ view around SNR\,G54.1+0.3 is shown on the right inset within the black box. The centroid of a known TeV source, VER\,J1930+188, measured by VERITAS, is shown with a yellow circle in the right inset. The centroid of the source measured by H.E.S.S. is shown with a pink circle in the inset. In the region around VER\,J1930+188, \textit{Fermi}-LAT data are best described with a model with two point sources whose locations are indicated with black diamond markers. The point source coinciding with the VERITAS source has a TS value of 23, while the other source has a TS value of 17. The four dark green crosses are the locations of 3FGL sources with dark green ellipses showing the 1$\sigma$ uncertainty of the location. The two blue x marks indicate the centroids of two HAWC sources in the region with 1$\sigma$ uncertainty of the position marked with blue circles. Four pulsars with a spin-down luminosity higher than $10^{35}$ $\textnormal{erg}$ $\textnormal{s}^{-1}$ are marked with red boxes. White contours are HAWC's significance contours of 5, 6, 7, and 8$\sigma$.} \label{Fig:G54region_skymap} 
\end{figure*}

\begin{figure}[t!]
 \centering
 \includegraphics[width=0.5\textwidth]{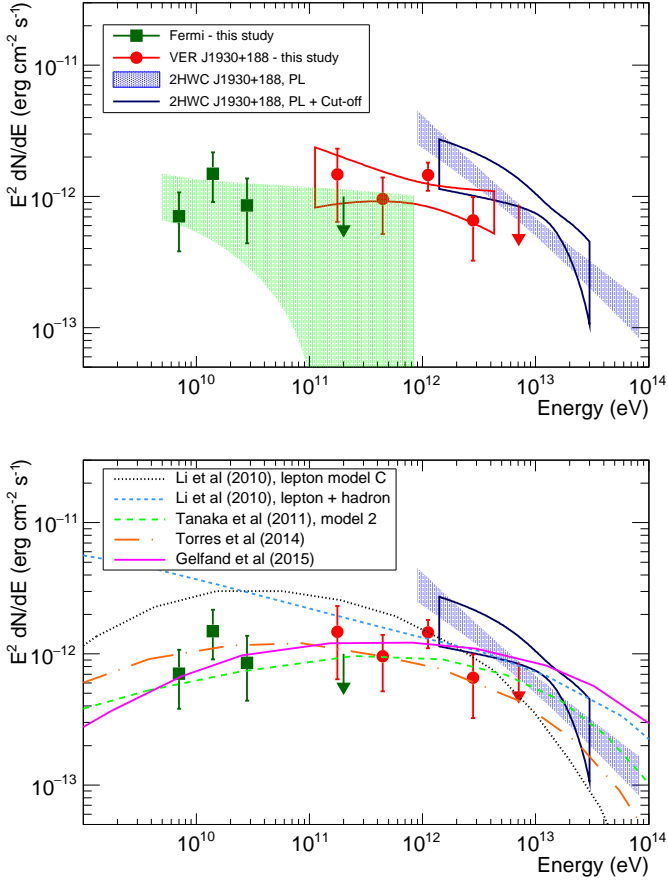}
 \caption{SEDs for sources in the SNR\,G54.1+0.3 region. The upper panel shows the SED of SNR\,G54.1+0.3, and the lower panel shows the existing models for SNR\,G54.1+0.3 together with the SED. The blue contours show the 1$\sigma$ confidence interval of HAWC's spectrum measurements for a single power law distribution (filled) and for a power law with a cutoff (empty). Red points and upper limits are the VERITAS measurements. Green colored rectangles and upper limits are the measurements of the newly detected \textit{Fermi} source in the SNR\,G54.1+0.3 region. The green filled area shows the 1$\sigma$ confidence interval estimated by using the statistical errors of the \textit{Fermi} spectral analysis. The black dotted line shows the best fit leptonic model from \cite{2010MNRAS.408L..80L}, and the light blue short dashed line shows the lepton-hadron model from \cite{2011ApJ...741...40T}. The orange broken dashed line and the magenta solid line show models from \cite{2014JHEAp...1...31T} and \cite{2015ApJ...807...30G}.}
 \label{Fig:G54Spectrum}
\end{figure}

\textit{Fermi}-LAT analysis of this region detected a point source coincident with VER\,J1930+188 with a TS value of 26. The centroid of the point source is RA, Dec (J2000) =$\textnormal{19}^{\textnormal{h}}\textnormal{30}^\textnormal{m}\textnormal{16.8}^\textnormal{s}$, $\textnormal{18}^\circ \textnormal{55}\arcmin\textnormal{48.0}\arcsec$ with an uncertainty of $1.8\arcmin$. 
As shown in Figure~\ref{Fig:G54Spectrum}, the flux measured by LAT is consistent with VERITAS measurements. The non-detection of SNR\,G54.1+0.3 in the 3FGL catalog indicates the possible existence of a low-energy spectral cutoff in the \textit{Fermi} energy range, but the source is too faint to measure the cutoff in this study. 

We also explored different possible spatial morphologies for this new \textit{Fermi}-LAT gamma-ray source. The results of these tests are recorded in Table \ref{Tbl:LATmorphology}. Here the figures of merit are the change in TS and the number of degrees of freedom, N$_{\rm dof}$, where TS is equal to twice the difference between the log likelihoods of the null hypothesis and the tested spatial model. We first tested an extended model assuming a uniformly illuminated disk-shaped source at the location of SNR\,G54.1+0.3 with an initial radius of $0.2\degree$. The best-fit extension was found to be a radius of $0.4\degree\pm0.1\degree$. The change in TS between the point source hypothesis and the tested spatial model is referred to as TS$_{\rm ext}$. The uniform disk gives TS$_{\rm ext}$ of 22 with one additional degree of freedom.   
Past LAT analyses and simulations of spatially extended sources find that a TS$_{\rm ext} \geq 16$ corresponds to a formal 4$\sigma$ significance~\citep{2012ApJ...756....5L}. Following \citet{2017arXiv170200476T}, a source is considered to be extended only if TS$_{\rm ext}$ $>$ TS$_{\rm 2pts}$, where TS$_{\rm 2pts}=2\ln(L_{\rm 2pts}/L_{\rm ps}$) is the improvement when adding a second point source, $L_{\rm 2 pts}$ is the likelihood of the model with two point sources, and $L_{\rm ps}$ is the likelihood of the model with a single point source. In the case of 2HWC\,J1930+188, the extended LAT source did not meet this criterion, and so was replaced by two point sources located at the best positions found by our localization algorithm. The centroids of the two point sources are plotted as black diamonds in Figure~\ref{Fig:G54region_skymap}. The additional point source is located 0.3\degree\ away from SNR\,G54.1+0.3 and has a TS of 17. The addition of the second source also caused the TS of the point source that is spatially coincident with SNR\,G54.1+0.3 to decrease from 26 to 23. In summary, there is not yet sufficient data to establish the spatial morphology of the LAT counterpart to SNR\,G54.1+0.3, and we therefore prefer a single LAT point source.
\tabletypesize{\normalsize}
\begin{deluxetable}{l ll}[ht!!]
\tablecaption{Spatial models for LAT analysis in the vicinity of SNR\,G54.1+0.3\label{Tbl:LATmorphology}}
\tablehead{
\colhead{Spatial Model} & \colhead{$\rm TS$} & \colhead{N$_{\rm dof}$}
}
\startdata
Null Hypothesis & \ldots & \ldots \\
Point Source & 26 & 4 \\
Uniform Disk & 48 & 5 \\
Two Point Sources & 55 & 8 \\
\enddata
\end{deluxetable}

With 16 hours of additional data taken in the 2015--2016 observing season, we studied a total of 46 hours of VERITAS exposure in this region. VER\,J1930+188 is a point-like source for VERITAS. The updated spectrum now extends down to 120~GeV with an index of $2.18 \pm 0.20_{\textnormal{stat}}$. This is in agreement with the previous result, $2.39 \pm 0.23_{\textnormal{stat}}$, within 1$\sigma$~\citep{2010ApJ...719L..69A}. The updated normalization value at 1~TeV is $( 6.6 \pm 1.3_{\textnormal{stat}})\times 10^{-13}$~$\textnormal{TeV}^{-1}\textnormal{cm}^{-2}\textnormal{s}^{-1}$. For energies higher than 4.9~TeV, the 99$\%$ upper limit is $1.07\times10^{-14}$ $\textnormal{cm}^{-2}$ $s^{-1}$ $\textnormal{TeV}^{-1}$ at 7.08~TeV assuming a spectral index of 2.18. 

The HAWC source 2HWC\,J1930+188, coincident with VER\,J1930+188, was detected in the point source search. The centroid of 2HWC\,J1930+188 is shown in Figure~\ref{Fig:G54region_skymap}, and agrees with both the VERITAS and \textit{Fermi} centroid positions. As shown in Figure~\ref{Fig:G54Spectrum}, the spectral index of the HAWC source, 2.74 $\pm$ 0.12$_{\textnormal{stat}}$, is softer than that measured by VERITAS. The significance of the difference is 2.4$\sigma$ considering the statistical errors, and 1.5$\sigma$ considering both systematic ($\sim$0.2) and statistical errors. Extrapolation of the HAWC spectrum to the VERITAS energy range yields an integrated flux that is seven times larger than the VERITAS flux. Although this is still in agreement with the VERITAS measurement within 2$\sigma$ statistical uncertainties, we tested whether the HAWC data favor a power law distribution with a cutoff. To reduce the number of degrees of freedom, we fixed the index of the power law with a cutoff scenario to the index value measured by VERITAS. The results are plotted in Figure~\ref{Fig:G54Spectrum}. 
As summarized in Table~\ref{Tbl:HAWCspectrumTestForSNRG54}, the HAWC result can be explained with either a single power law or a power law with a cutoff. The extrapolation of the power law with a cutoff to VERITAS energies produces an integral flux that is only $\sim$50\% larger than the VERITAS flux, within the 1$\sigma$ statistical error, providing better agreement. While all three measurements were estimated for a point-like source, HAWC would estimate flux from a larger area than VERITAS due to their larger PSF. Because HAWC modeled a single source in the likelihood analysis for this study, the flux estimation may be influenced by emission from other sources in the region.

\begin{deluxetable}{c cc} [t!]
\tabletypesize{\small}
\tablecaption{Spectrum for 2HWC\,J1930+188 with two different spectral models\label{Tbl:HAWCspectrumTestForSNRG54}}
\tablehead{
\colhead{} &\colhead{PL} & \colhead{PL+cutoff} \\
\colhead{$E_\mathrm{12.5}$--$E_\mathrm{87.5}$ (TeV)} &\colhead{0.9--86} & \colhead{1.4--30} \\
}
\startdata
Index & 2.74$\pm$0.12 & 2.18 (fixed)  \\
Norm at 7 TeV & \multirow{2}{*}{9.8$\pm$1.5} & \multirow{2}{*}{19.6$\pm$9.0}\\
(10$^{-15}$ TeV$^{-1}$cm$^{-2}$s$^{-1}$) & &  \\
Cutoff energy & \multirow{2}{*}{N/A} & \multirow{2}{*}{21$\pm$15} \\
(TeV) & & \\
TS & 54 & 52 \\
\enddata
\end{deluxetable}

The likely astrophysical counterpart for both the newly detected \textit{Fermi} point source, VER\,J1930+188, and 2HWC\,J1930+188 is SNR\,G54.1+0.3, a PWN at a distance of $\sim$$\textnormal{6.5}\U{kpc}$ hosting a young, energetic pulsar, PSR\,J1930+1852, with a spin-down luminosity of $1.2\times10^{37} \textnormal{erg}$ $\textnormal{s}^{-1}$ and a characteristic age of $2900\U{years}$~\citep{2002ApJ...574L..71C}. The pulsar powers a PWN, which is observed in radio and X-rays. The gamma-ray emission from the PWN can be explained as resulting from inverse Compton scattering of electrons accelerated at the PWN termination shock on ambient photon fields. 

In the X-ray band, a torus structure with a size of 5.7$\arcsec$ by 3.7$\arcsec$ is detected with a jet~\citep{2002ApJ...568L..49L,2010ApJ...710..309T}, while the diffuse emission covers a larger area with a size of 2.0$\arcmin$ by 1.3$\arcmin$~\citep{2010ApJ...710..309T}. The extension of the diffuse emission is similar in the radio~\citep{2010ApJ...709.1125L} and X-ray bands~\citep{2010ApJ...710..309T}. Even including the diffuse emission, the PWN is a point-like source for LAT, VERITAS and HAWC. 

The SED of SNR\,G54.1+0.3 has been subjected to detailed study, and a number of authors have attempted to construct models of the system based on the observed emission from the radio band up to the gamma-ray band~\citep{2010MNRAS.408L..80L,2011ApJ...741...40T,2014JHEAp...1...31T, 2015ApJ...807...30G}. The lower panel of Figure~\ref{Fig:G54Spectrum} shows the existing models together with the gamma-ray SED of the PWN emission. All of the models assume a spatially uniform magnetic field strength and particle density, and include the effect of the time-dependent evolution of the PWN assuming a broken power law distribution of electron energies. Although their assumptions about the environment of the PWN and particle diffusion are slightly different, the estimated gamma-ray emission is similar for all models except in the case of the lepton-hadron model suggested by \cite{2010MNRAS.408L..80L}. \cite{2010MNRAS.408L..80L} argued for the lepton-hadron model because the low magnetic field strength of $\sim$10~$\mu$G required for the pure leptonic model is inconsistent with the $\sim$38~$\mu$G estimated by \cite{2010ApJ...709.1125L} based on the radio luminosity. \cite{2010MNRAS.408L..80L} also commented that the lepton-hadron model reproduced the reported VERITAS measurement~\citep{2010ApJ...719L..69A} better than the pure leptonic model. Some other authors~\citep{2011ApJ...741...40T,2014JHEAp...1...31T} pointed out that the value derived from the observation is based on an assumption that the energy of the pulsar wind is equally divided between the magnetic field and the particle energies. 
However, all other models favor a very small contribution of the wind energy to the magnetic field, ranging from 0.04$\%$ to 0.5$\%$, indicating that the PWN is a particle-dominated nebula. The lepton-hadron model is also disfavored by the \textit{Fermi}-LAT flux measured in this study, as shown in Figure~\ref{Fig:G54Spectrum}. The gamma-ray emission from 3~GeV up to 100~TeV generally agrees well with the other models~\citep{2011ApJ...741...40T,2014JHEAp...1...31T, 2015ApJ...807...30G}. However the soft spectral index of HAWC at energies above 1~TeV under the single PL assumption, or the low cutoff energy of 21$\pm$15~TeV under the PL with exponential cutoff assumption, indicate that the maximum electron energy may be smaller than the 90\% confidence interval of 0.96--2700~PeV reported by \cite{2015ApJ...807...30G}.  

\subsubsection{2HWC\,J1928+177}
The other HAWC source in the region is 2HWC\,J1928+177. HAWC reported a similar value of flux and index for this source as for 2HWC\,J1930+188. HAWC analysis shows that 2HWC\,J1928+177 is brighter than 2HWC\,J1930+188 for energies higher than 10~TeV~\citep{2017arXiv170803137L}. However, VERITAS did not detect emission from this source with either the point source search or the extended source search. The angular distance between 2HWC\,J1930+188 and 2HWC\,J1928+177 is $1.18\degree$, which is larger than the PSF of HAWC for energies larger than 1~TeV. Since 2HWC\,J1930+188 is point-like for TeV instruments, it is therefore likely that any contamination from it would result in only a slight overestimation of the flux from 2HWC\,J1928+177 as measured by HAWC. The lack of a VERITAS detection indicates that the HAWC source has a larger angular extent than the radius of $0.1\degree$ at the confidence level of 98$\%$ and the radius of $0.23\degree$ at the confidence level of 82$\%$.

\begin{figure}[t!]
 \centering
 \includegraphics[width=0.5\textwidth]{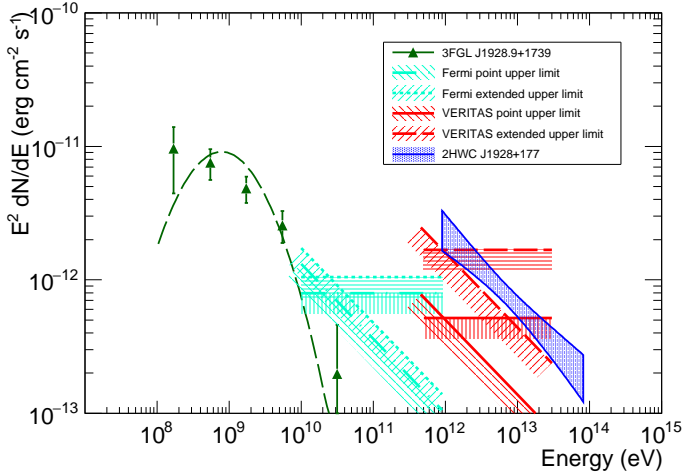}
 \caption{SED of 2HWC\,J1928+177. Blue filled area shows the $1\sigma$ confidence interval of HAWC's spectrum measurements. The solid red lines show the VERITAS upper limits from the point source search, assuming either the spectral index measured by HAWC or a spectral index of 2. The dashed red lines show the VERITAS upper limits from the extended source search for the same spectral indices. The corresponding \textit{Fermi}-LAT upper limits are shown as teal broken dashed lines for the point source search and teal dashed lines for the point source search and teal dashed lines for th extended source search. The green triangle points and dashed line show the flux of the nearby 3FGL source, 3FGL\,J1928.9+1739.} 
\label{Fig:Spectrum_2HWCJ1928}
\end{figure}

The position of a nearby \textit{Fermi} source, 3FGL\,J1928.9+1739, coincides with the position of 2HWC\,J1928+177 within $2\sigma$. Both sources are coincident with PSR\,J1928+1746, which at an age of 8.2$\times\textnormal{10}^4$~years and a spin-down luminosity of $\textnormal{1.6}\times\textnormal{10}^{36}$~$\textnormal{erg}$ $\textnormal{s}^{-1}$ is both older and less energetic than PSR\,J1930+1852~\citep{2006ApJ...637..446C}. The flux of 3FGL\,J1928.9+1739 follows a log parabola shape, and its extrapolation to energies larger than 1~TeV lies far below the flux of 2HWC\,J1928+177, as shown in Figure~\ref{Fig:Spectrum_2HWCJ1928}. If the nature of both sources is tied to PSR\,J1928+1746, then it is possible that 3FGL\,J1928.9+1739 corresponds to the pulsed emission of PSR\,J1928+1746, while 2HWC\,J1928+177 may originate from a PWN. However, no pulsation is reported for 3FGL\,J1928.9+1749. Also, no PWN has been observed for PSR\,J1928+1746 in any other wavelength. 

\subsubsection{Other gamma-ray emission in the region}
Although there are only two HAWC sources reported in this region, the extension of HAWC's $\textnormal{5}\sigma$ contours covers a larger area than these two sources, as shown in Figure~\ref{Fig:G54region_skymap}. It is possible that there are other weak, and possibly extended, TeV emitting sources yet to be identified in this region. 

HAWC's $\textnormal{5}\sigma$ contours contain four 3FGL sources. These include 3FGL\,J1928.9+1739, which has been described in the previous section; 3FGL J1932.2+1916, a LAT pulsar; and two unassociated LAT sources, 3FGL\,J1928.4+1838 and 3FGL\,J1925.4+1727. 3FGL J1925.4+1727 is located nearby a young pulsar, PSR\,J1925+1720, which has a spin-down luminosity of $10^{36}$ $\textnormal{erg}$ $\textnormal{s}^{-1}$~\citep{2017ApJ...834..137L}. Similar to 3FGL\,J1928.9+1739, the extrapolation of the SEDs of these two unassociated LAT sources to energies higher than 1~TeV yields less than 1$\%$ of the Crab PWN flux. Thus, likely this source does not directly correspond to what HAWC measured in the region. 

Future observations with longer exposure from HAWC and follow-up from IACTs will be necessary to study the nature of VHE gamma-ray emission in the region and the connections with these unassociated \textit{Fermi}-LAT sources. 

\subsection{DA\,495 region}\label{Sec:DA495region}
\begin{figure}[ht!]
  \centering
  \includegraphics[width=0.5\textwidth]{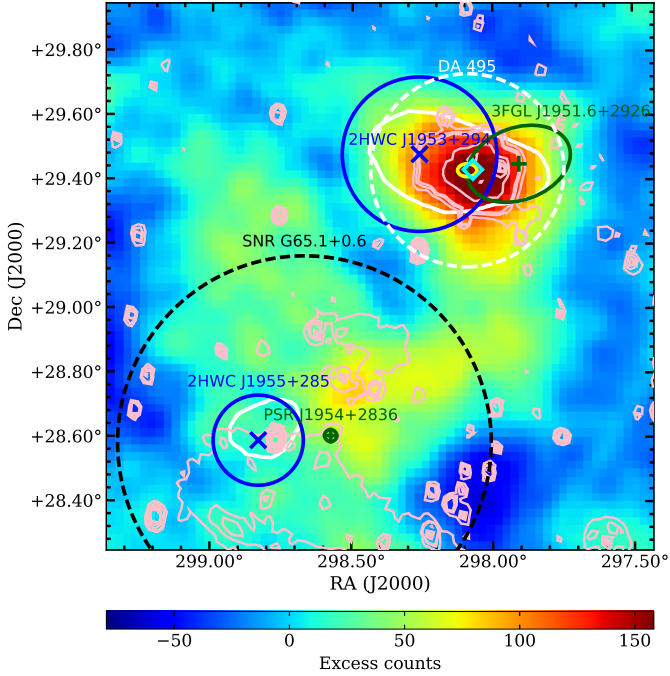}
\caption{VERITAS gamma-ray counts map of the PWN\,DA\,495 and the surrounding region with extended source search cuts, $\theta<$0.3$^{\circ}$. The region used for the spectral analysis with VERITAS data is shown with a white dashed circle. The yellow circle is the centroid measured by VERITAS. Two dark green crosses and ellipses are the location and 1$\sigma$ uncertainty of the location of 3FGL sources. Blue x marks indicate the centroids of two HAWC sources in the region with blue circles showing 1$\sigma$ uncertainty. White contours are HAWC's significance contours of 5$\sigma$. The cyan diamond is the location of an X-ray compact source, 1WGA J1952.2+2925~\citep{2004ApJ...610L.101A}. Light pink contours show the radio contours around PWN\,DA\,495 measured by the Canadian Galactic Plane Survey in the 1.42 GHz band~\citep{2003AJ....125.3145T}. The extension of radio emission from SNR\,G65.1+0.6 is marked with a dashed black line.} \label{Fig:DA495region_skymap}
\end{figure}

The second area we discuss in detail is a region around PWN\,DA\,495 (SNR\,G065.7+01.2). HAWC detected two point-like sources in this region: 2HWC\,J1953+294 and 2HWC\,J1955+285.

Analysis of \textit{Fermi}-LAT data for the energy range from 10~GeV to 900~GeV did not detect gamma-ray emission in either the point source search or the extended source search. 

\subsubsection{PWN DA\,495}
After 37 hours of observation, VERITAS reported a confirmation of weak gamma-ray emission coincident with 2HWC\,J1953+294~\citep{2016arXiv160902881H} with an extended source analysis ($\theta$$<$$0.3\degree$). After this initial report, VERITAS continued observing the source, and accumulated a total of 72 hours of data on the field of view by summer 2017. With a maximum significance of 5.2$\sigma$, VERITAS detected emission nearby 2HWC\,J1953+294. The emission observed by VERITAS is centered at RA $19^h52^m15^s$ $\pm$ $9^s_{\textnormal{stat}}$, Dec $29\degree 23{\arcmin}$ $\pm$ $01{\arcmin}_{\textnormal{stat}}$, assuming that the spatial distribution of the emission follows a simple 2D-Gaussian distribution; hence, we assign the name VER\,J1952+293. The best-fit sigma value of the 2D-Gaussian is $0.14^{\circ}$ $\pm$ $0.02^{\circ}_{\textnormal{stat}}$. The distribution of gamma-ray emission in the region observed by VERITAS is shown in Figure~\ref{Fig:DA495region_skymap}.

\begin{figure}[t!]
  \centering
  \includegraphics[width=0.5\textwidth]{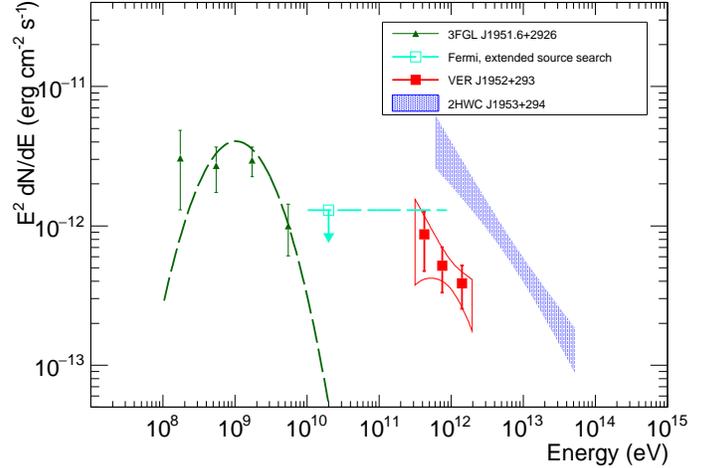}
\caption{\textit{Fermi}-LAT, VERITAS, and HAWC flux measurements in the vicinity of the PWN\,DA\,495. Green triangle points and dashed line show the flux of nearby 3FGL sources, 3FGL\,J1951.6+2926. The teal colored rectangular points are the upper limits of \textit{Fermi}-LAT data assuming the extension of the source to be 25$\arcmin$. Blue filled regions are HAWC's flux measurements. Red circles show the upper limits from VERITAS and red squares are the flux measurement from VERITAS with $\theta$$<$0.3$^{\circ}$.} \label{Fig:DA495region_spectrum}
\end{figure}

The upper panel of Figure~\ref{Fig:DA495region_spectrum} shows the SEDs of gamma-ray emission measured around 2HWC\,J1953+294. The VERITAS flux was calculated by integrating gamma rays within a $0.3^{\circ}$ radius around the centroid, while the HAWC flux was estimated based on a point-like assumption. The flux measured by VERITAS is well described by a power law distribution with an index of $2.65\pm0.49_{\textnormal{stat}}$, which is in good agreement with the index measured by HAWC, $2.78\pm 0.15_{\textnormal{stat}}$. But, the flux normalization value at 1~TeV measured by VERITAS is $(\textnormal{2.84}\pm\textnormal{0.54}_{\textnormal{stat}}) \times\textnormal{10}^{-13}$~$\textnormal{cm}^{-2} \textnormal{s}^{-1} \textnormal{TeV}^{-1}$, about seven times lower than the extrapolated flux value of HAWC's measurement to 1~TeV, $1.86\times\textnormal{10}^{-12}$~$ \textnormal{cm}^{-2} \textnormal{s}^{-1} \textnormal{TeV}^{-1}$. The difference between the two measurements is significant at the level of 2.4$\sigma$ when considering only statistical errors. Because VERITAS's analysis is using $0.3\degree$ as an integration radius while HAWC assumes it to be a point-like source, potentially the flux disagreement can be larger. Unlike the discrepancy shown in SNR\,G54.1+0.3, the index measurements by the two instruments are in good agreement and the energy ranges of both measurements are very similar. This suggests that a change of spectral index may not be the reason for the discrepancy of the flux estimations. 

We checked the details of both VERITAS and HAWC analyses for potential causes of the discrepancy. For the background estimation, VERITAS excluded events falling within 0.3$^{\circ}$ of 2HWC\,J1955+285 and two bright stars in the field of view. The same exclusion radius was used for the pulsars PSR\,J1954+2836 and PSR\,J1958+2846, while an exclusion region of radius 0.34$^{\circ}$ was used for 2HWC\,J1953+294 in order to cover the extension of the radio PWN. The background distribution after these exclusions was reasonable. If there is diffuse emission covering a very large area and weak enough not to be detected by the extended analysis of HAWC, this kind of discrepancy is possible. This is because the standard analysis of VERITAS obtains its background regions from the same field of view while HAWC's flux estimation would include both source and diffuse emission.  For the HAWC analysis, we re-estimated the HAWC flux after adding a uniform diffuse source to the model for the emission, which reduced HAWC's flux for 2HWC\,J1953+294 only by 10--15\%. The flux reported in the 2HWC catalog can be also overestimated if there is a nearby source, because HAWC's analysis performed for the 2HWC catalog assumes a single source for their likelihood analysis. 
We re-estimated the flux of 2HWC\,J1953+294 after removing the contribution from the nearby source, 2HWC\,J1955+285, assuming it to be a point-like source. The result shows only $\sim$$\textnormal{3}\%$ smaller flux for 2HWC\,J1953+294 compared to what was reported in the 2HWC catalog.
It is possible that the nearby 2HWC \,J1955+285 is extended, in which case the flux of DA\,495 reported by HAWC may be overestimated due to contamination from this source. A scenario in which 2HWC J1955+285 is extended could also better explain the VERITAS non-detection of this source. 

Drawing firm conclusions about the discrepancy between the VERITAS and HAWC observations in this field of view is challenging because of the relatively weak signals (TS of 25--30) of the sources reported in the 2HWC catalog. Further detailed study with larger HAWC exposure and advanced analysis including multi-source likelihood analysis will be necessary to understand the discrepancy.

We did not detect gamma-ray emission from 2HWC\,J1953+294 with the \textit{Fermi}-LAT analysis. The upper limit was obtained by assuming the source extension to be similar to the extension of radio emission from DA\,495, which is $\sim$25$\arcmin$~\citep{2008ApJ...687..516K}. The upper limit at the 99$\%$ confidence level is $\textnormal{8.17}\times\textnormal{10}^{-11} \textnormal{cm}^{-2}\textnormal{s}^{-1}$ with an assumption of spectral index of 2.78, and $\textnormal{8.00}\times\textnormal{10}^{-11} \textnormal{cm}^{-2}\textnormal{s}^{-1}$ with an assumption of spectral index of 2. Assuming that the spectral index measured by HAWC will not change down to 10~GeV, the upper limit we calculated disagrees with HAWC's flux estimation at a confidence level of 85$\%$. 

The likely counterpart of 2HWC\,J1953+294 and VER\,J1952+293 is the PWN\,DA\,495. As shown in Figure~\ref{Fig:DA495region_skymap}, the emission seen by VERITAS overlaps with the radio contours of DA\,495, an X-ray compact source, 3FGL\,J1951.61+2926 and 2HWC\,J1953+294. DA\,495 is seen as an extended, central concentration of emission in the radio band. X-ray observations by \textit{ROSAT} and \textit{ASCA} revealed a compact central object, 1WGA\,J1952.2+2925, surrounded by an extended nonthermal X-ray source~\citep{2004ApJ...610L.101A}. The implied blackbody temperature and luminosity, measured by \textit{Chandra}, suggest that the central object is an isolated neutron star. Together with the extended emission surrounding the compact object, this confirms the PWN interpretation of the source~\citep{2004ApJ...610L.101A,2008ApJ...687..505A}. \cite{2008ApJ...687..516K} suggested that DA\,495 may be an aging PWN with an age of $\sim$20,000 yr, based on the low-energy break measured in the radio band. Non-detection of an SNR shell suggests that the supernova exploded in a very low density environment. The distance to DA\,495 is estimated to be $1.0\pm 0.4$ kpc based on H I absorption measurements~\citep{2008ApJ...687..516K}. The extension of the PWN is $25\arcmin$~\citep{2008ApJ...687..516K} in radio and $\sim$$40\arcsec$ in X-ray~\citep{2008ApJ...687..505A}. The detected TeV extension by VERITAS matches well with the radio extension. 

There is a $\textit{Fermi}$-LAT source, 3FGL\,J1951.61+2926 coincident with DA\,495. The extrapolated flux of this source to the HAWC energy range is much lower than the flux measured by HAWC, as shown in Figure~\ref{Fig:DA495region_spectrum}. \cite{2015MNRAS.453.2241K} suggested that 3FGL\,J1951.61+2926 is likely associated with the central pulsar of DA\,495, although no evidence for pulsations has been identified. 

\begin{figure}[t!]
  \centering
  \includegraphics[width=0.5\textwidth]{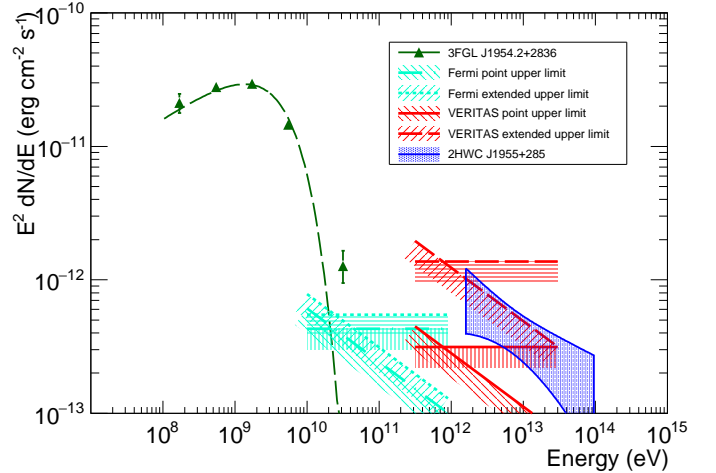}
\caption{SED of 2HWC\,J1955+285. The description of the blue filled region, red lines, and teal lines are the same as in Figure~\ref{Fig:Spectrum_2HWCJ1928}. The green triangle points and dashed line show the flux of the nearby 3FGL source, 3FGL\,J1954.2+2836.} 
\label{Fig:2HWC_J1955+285}
\end{figure}

\subsubsection{2HWC\,J1955+285}
The second HAWC source in the region, 2HWC\,J1955+285, is 1$\degree$ away from 2HWC\,J1953+294. There is a nearby radio-quiet gamma-ray pulsar, PSR\,J1954+2836, detected by \textit{Fermi}-LAT (3FGL\,J1954.2+2836). The positions of this pulsar and 2HWC\,J1955+285 agree within $2\sigma$. But, similar to DA\,495, the extrapolation of PSR\,J1954+2836 to higher energies lies far below HAWC's flux measurement as shown in Figure~\ref{Fig:2HWC_J1955+285}. 

These sources are within the extent of SNR\,G65.1+0.6, a very faint shell-type SNR detected in the radio band~\citep{1990A&A...232..207L,2006A&A...457.1081K}. \citet{2006A&A...455.1053T} reported the distance to the SNR to be 8.7--10.1 kpc and the average diameter of the SNR to be 70$\arcmin$. The large size and diffusive appearance of the SNR suggests that it likely exploded in a low density environment~\citep{2006A&A...457.1081K,2006A&A...455.1053T}. The size of the SNR is larger than the PSF of HAWC. However, even though 2HWC\,J1955+285 is found in the HAWC point source search, the 2HWC analysis cannot rule out the possibility of the source being extended. As mentioned in the previous section, the measured flux discrepancy between VERITAS and HAWC on DA~\,495 can be explained as an overestimation of the 2HWC flux due to contamination from 2HWC\,J1955+285 if this source has a large extension. Further studies with additional HAWC exposure will be needed to clarify the connection between 2HWC\,J1955+285 and SNR\,G65.1+0.6. 

Figure~\ref{Fig:DA495region_skymap} shows that VERITAS sees a region of excess gamma-ray counts around 2HWC\,J1955+285. The maximum pre-trial significance in this region is 3.5$\sigma$ offset by 0.35$\degree$ from the position of the HAWC source. With the current data set, it is unclear whether this is a weak source or simply a statistical fluctuation. 

\section{Discussion}
\subsection{VERITAS follow-up studies of unassociated 2HWC sources }
As shown in Figure~\ref{Fig:EnergyRangeComparisons}, there is large overlap between the energy range covered by HAWC and VERITAS for most of the sources selected for this study. Since the VERITAS sensitivity for a point source is better than that of HAWC, VERITAS should be able to detect a small-sized (radius $<$ 0.5$^{\circ}$) source with a substantially shorter exposure time. The angular resolution of VERITAS is also better than that of HAWC, so a VERITAS measurement should provide additional insight into the source morphology and extension.

New HAWC sources generally have low TS values compared to the 2HWC sources associated with known TeV-emitting sources. Among the fourteen sources we selected for this follow-up study, eleven sources have a TS value less than 36 except for 2HWC\,J1852+013$^{\ast}$, 2HWC\,J1928+177, and 2HWC\,J2006+341. The average spectral index of the selected sources is 2.6, with individual indices ranging from 1.5 to 3.3. 

Among the fourteen selected new sources, four were detected by an extended source search with HAWC. These sources are 2HWC\,J0700+143, 2HWC\,J0819+157, 2HWC\,J1040+308, and 2HWC\,J1949+244. Of these, 2HWC\,J0700+143 and 2HWC\,J1949+244 were detected by the search for sources with an extension of 1$\degree$, while 2HWC\,J0819+157 and 2HWC\,J1040+308 were detected by the search for sources with an extension of 0.5$\degree$. The exposure of VERITAS on these sources is relatively small (1.8$\sim$5.8 hours), and the upper limits are not strongly constraining. 

The other ten sources were found by a search for point-like sources with HAWC. Nine of the ten sources were not detected by VERITAS, and we find that the 99$\%$ flux upper limits from VERITAS (assuming a point-source hypothesis) are lower than the expected flux obtained from the best-fit spectra provided by HAWC. Treating the uncertainties in the HAWC fluxes as Gaussian, and considering the statistical errors only, we can exclude six sources--2HWC\,J1852+013$^{\ast}$, 2HWC\,J1902+048$^{\ast}$, 2HWC\,J1928+177, 2HWC\,J1938+238, 2HWC\,J2006+341, 2HWC\,J2024+417$^{\ast}$-- as being point sources with the same power law energy distribution as measured by HAWC with 95$\%$ confidence level. It is possible to explain this disagreement with a changing spectral shape, as we have demonstrated with SNR\,G54.1+0.3, if the source is indeed point-like to VERITAS and HAWC. 

In the extended source analysis by VERITAS, the upper limits are less constraining. With angular cuts of 0.23$\degree$, the upper limits measured by VERITAS agree with the flux estimated from HAWC for all but three sources: 2HWC\,J1852+013$^{\ast}$, 2HWC\,J1902+048$^{\ast}$, and 2HWC\,J1907+084$^{\ast}$. The discrepancy between the VERITAS and HAWC measurements is especially large for 2HWC\,J1852+013$^{\ast}$ and 2HWC\,J1902+048$^{\ast}$. The measurements for these two sources disagree at a confidence level of greater than 95$\%$. Both of these sources have relatively large VERITAS exposures ($>$10 hours). To satisfy both the VERITAS upper limit and the measured HAWC flux, the source extension must be larger than a radius of 0.23$\degree$. 

\begin{figure*}[htb]
  \centering
  \includegraphics[width=\textwidth]{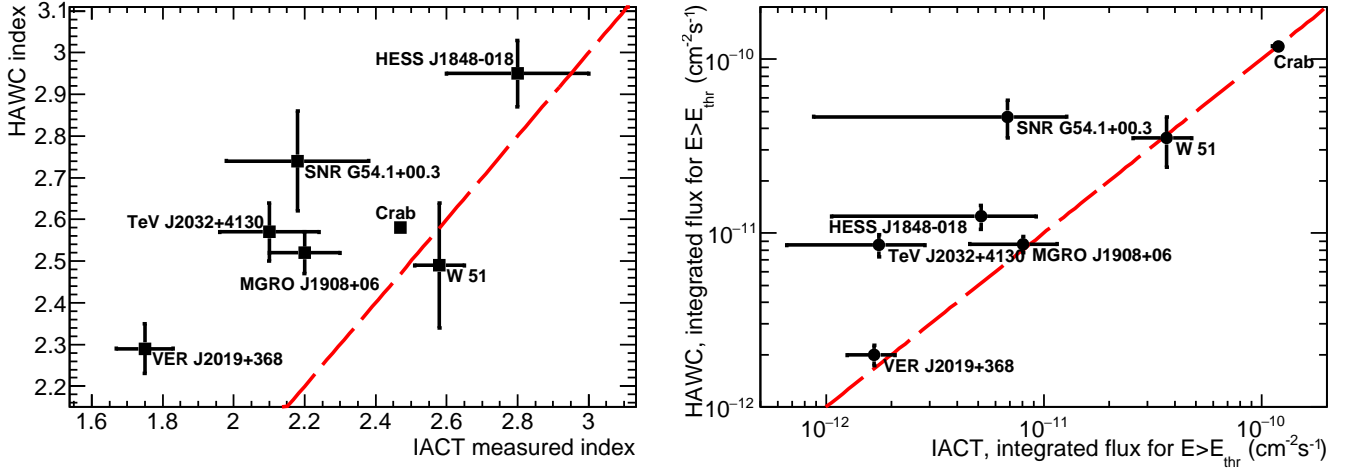}
\caption{Spectral index (left) and integral flux (right) comparisons between HAWC and IACTs for known TeV sources. The energy range used to calculate the integrated flux for each source was chosen to start at the threshold value for the IACT measurement and end at 30 TeV. Red lines are for case that the two measurements exactly match.} \label{Fig:FluxComparisons_for_known_sources} 
\end{figure*}

\subsection{Comparison between IACT measurements and 2HWC source properties for TeV sources previously detected by IACTs}

We compared the flux measurements by VERITAS, MAGIC, H.E.S.S., and HAWC for sources previously detected by IACTs to check whether there are systematic differences in fluxes measured by IACTs and HAWC, which may explain the constraining upper limits measured in this study. We selected only isolated HAWC sources if each source is coincident with a single TeV source previously detected by IACTs. We required the distance between the HAWC source and the selected TeV source detected by IACTs to be smaller than 0.2$\degree$, and the TeV source to have published spectral measurements. A total of seven sources were selected---the Crab Nebula, HESS\,J1848-018~\citep{2008AIPC.1085..372C}, MGRO\,J1908+06~\citep{2014ApJ...787..166A}, W\,51~\citep{2012A&A...541A..13A}, SNR\,G54.1+0.3~\citep{2010ApJ...719L..69A}, VER\,J2019+368~\citep{2014ApJ...788...78A}, and TeV\,J2032+4130~\citep{2008ApJ...675L..25A,2014ApJ...783...16A}. Except for the Crab Nebula and SNR\,G54.1+0.3, all sources were observed as extended sources by the IACTs. Following the procedure described in Section~\ref{Sec:Sources-not-VERITAS}, we used spectral information provided in the HAWC point source search to calculate the integral flux of HAWC in the energy range measured by IACTs. Then, these values were compared to the integral fluxes measured by IACTs. The energy threshold ($E_{thr}$) for each source reported by IACTs varies from 75~GeV (W51) up to 1~TeV (MGRO\,J2019+368).

Figure~\ref{Fig:FluxComparisons_for_known_sources} shows the results. As visible in the left panel, HAWC generally sees softer spectra than IACTs for these sources. However, as shown in the right panel, integral flux measurements by IACTs and HAWC agree well for most of the sources, with the largest discrepancies appearing for SNR\,G54.1+0.3 and TeV\,J2032+4130. 
We discussed the flux difference for SNR\,G54.1+0.3 in Section~\ref{Sec:SNRG54.1}. TeV\,J2032+4130 is located inside a large GeV emission region measured by \textit{Fermi}-LAT, also known as the Cygnus cocoon~\citep{2011Sci...334.1103A}. Given the large PSF of HAWC, it is likely the diffuse emission from the cocoon increased the flux estimated by HAWC even with their point source assumption for the flux of 2HWC\,J2031+415. 

By comparing the fluxes of known, isolated sources detected by both IACTs and HAWC, we conclude that there is no clear and large systematic difference in the fluxes estimated by these instruments that could explain the non-detection of the new HAWC sources by VERITAS. A change in spectral shape, source extension, or an overestimation of the HAWC flux due to additional diffuse emission in the source vicinity is likely the cause. 

\subsection{\textit{Fermi}-LAT follow-up studies of unassociated 2HWC sources}

Extrapolation of HAWC spectra to the \textit{Fermi}-LAT energy range results in large uncertainties, so the flux estimates from HAWC and upper limits measured by \textit{Fermi}-LAT agree within 1--2$\sigma$ for most of the sources for both the point-like and extended source searches. The most clear disagreement between two measurements is for 2HWC\,J1852+013$^{\ast}$ as shown in the appendix; the measurements are discrepant at a confidence level of 94$\%$ for both the point-source search and the extended source search . All of the upper limits measured by \textit{Fermi}-LAT, except for that of 2HWC\,J0819+157, are lower than the extrapolation of the HAWC spectrum. Combined with the VERITAS results, this suggests that there are likely spectral shape changes between the \textit{Fermi}-LAT energy range and the HAWC energy range. This also could explain why there were no 3FHL sources coincident with the selected HAWC sources. 

\begin{figure}[t!]
  \centering
  \includegraphics[width=0.5\textwidth]{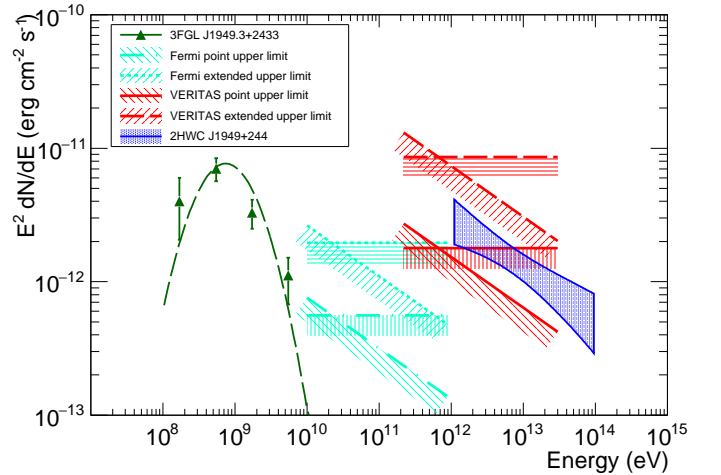}
\caption{SED of 2HWC\,J1949+244. The description of the blue filled region, red lines, and teal lines are the same as in Figure~\ref{Fig:Spectrum_2HWCJ1928}. The green triangle points and dashed line show the flux of the nearby 3FGL source, 3FGL\,J1949.3+2433. } 
\label{Fig:2HWCJ1949+244}
\end{figure}

Comparing the 3FGL catalog with the HAWC catalog, we found that there are 3FGL sources in the vicinity of four HAWC sources--3FGL\,J1928.9+1739, 3FGL\,J1949.3+2433, 3FGL\,J1951.6+2926, and 3FGL\,J1954.2+2836. Three of these sources were discussed in the previous sections. The remaining source, 3FGL\,J1949.3+2433, is in the vicinity of 2HWC\,J1949+244. The flux measurements in the vicinity of 2HWC\,J1949+244 are shown in Figure~\ref{Fig:2HWCJ1949+244}. All of these 3FGL sources have SEDs that decrease sharply above energies higher than a few GeV, and the extrapolations of the SEDs to the HAWC energy range produce fluxes that are much lower than the HAWC measurements. Whether there is any connection between these two sets of measurements, such as pulsar emission measured by \textit{Fermi}-LAT and PWN emission measured by HAWC, can be verified with further observations with IACTs and other multiwavelength observations.

\section{Conclusion}
Using VERITAS and \textit{Fermi}-LAT, we searched for IACT and GeV gamma-ray counterparts to  fourteen out of nineteen new HAWC sources without clear TeV associations. VERITAS detected one weak source coincident with PWN DA\,495. The flux of DA\,495 measured by VERITAS is about seven times lower than HAWC's measurement while both measurements agree on the spectral index. \textit{Fermi}-LAT did not see gamma-ray emission for the selected fourteen sources for either point or extended source searches. \textit{Fermi}-LAT did detect point-like emission from SNR\,G54.1+0.3, a PWN detected by both VERITAS and HAWC. The combined SED of the three instruments on SNR\,G54.1+0.3 covers a wide range of the inverse Compton peak of the PWN, providing a good data set for future modeling. 

Upper limits measured by VERITAS are lower than expected from HAWC's measurement for nine sources. Among these, non-detections by VERITAS exclude a point-source hypothesis for six sources with a confidence level of 95$\%$. The discrepancy could be resolved if the sources are extended, or if there is a spectral change in the energy range between VERITAS and HAWC. For 2HWC\,J1852+013$^{\ast}$ and 2HWC\,J1902+048$^{\ast}$, the extension of the source should be larger than 0.23$\degree$ to satisfy all of the measurements. These numbers are based on a comparison between the upper limits of VERITAS and the flux estimation of HAWC. However, it is possible that the HAWC flux is overestimated for some of the sources, since the flux estimation has been made with a single point source model for the likelihood analysis without accounting for nearby sources. Unaccounted weak diffuse emission over a very large area would also cause an overestimation of the flux. While \textit{Fermi}-LAT will accumulate more exposure time, a future IACT like the Cherenkov Telescope Array (CTA) should be able to detect the sources with its larger field of view and improved sensitivity. A combined analysis with \textit{Fermi}-LAT, CTA and HAWC will provide detailed gamma-ray data to study the nature of these new VHE sources.

\acknowledgements
The VERITAS Collaboration is supported by grants from the U.S. Department of Energy Office of Science, the U.S. National Science Foundation and the Smithsonian Institution, and by NSERC in Canada. We acknowledge the excellent work of the technical support staff at the Fred Lawrence Whipple Observatory and at the collaborating institutions in the construction and operation of the instrument. 

The \textit{Fermi}-LAT Collaboration acknowledges generous ongoing support from a number of agencies and institutes that have supported both the development and the operation of the LAT as well as scientific data analysis. These include the National Aeronautics and Space Administration and the Department of Energy in the United States, the Commissariat \`a l'Energie Atomique and the Centre National de la Recherche Scientifique / Institut National de Physique Nucl\'eaire et de Physique des Particules in France, the Agenzia Spaziale Italiana and the Istituto Nazionale di Fisica Nucleare in Italy, the Ministry of Education, Culture, Sports, Science and Technology (MEXT), High Energy Accelerator Research Organization (KEK) and Japan Aerospace Exploration Agency (JAXA) in Japan, and the K.~A.~Wallenberg Foundation, the Swedish Research Council and the Swedish National Space Board in Sweden. This work performed in part under DOE Contract DE-AC02-76SF00515.

The HAWC Collaboration acknowledges support from: the US National Science Foundation (NSF); the US Department of Energy Office of High-Energy Physics; the Laboratory Directed Research and Development (LDRD) program of Los Alamos National Laboratory; Consejo Nacional de Ciencia y Tecnolog\'{\i}a (CONACyT), M{\'e}xico (grants 271051, 232656, 260378, 179588, 239762, 254964, 271737, 258865, 243290, 132197), Laboratorio Nacional HAWC de rayos gamma; L'OREAL Fellowship for Women in Science 2014; Red HAWC, M{\'e}xico; DGAPA-UNAM (grants IG100317, IN111315, IN111716-3, IA102715, 109916, IA102917); VIEP-BUAP; PIFI 2012, 2013, PROFOCIE 2014, 2015;the University of Wisconsin Alumni Research Foundation; the Institute of Geophysics, Planetary Physics, and Signatures at Los Alamos National Laboratory; Polish Science Centre grant DEC-2014/13/B/ST9/945; Coordinaci{\'o}n de la Investigaci{\'o}n Cient\'{\i}fica de la Universidad Michoacana. Thanks to Luciano D\'{\i}az and Eduardo Murrieta for technical support.

\software{Fermi Science Tools (v10r01p01), fermipy (v0.13;~\cite{2017arXiv170709551W}), APLpy(v1.1.1;~\cite{2012ascl.soft08017R}), ROOT(v5.34/36;~\cite{2011CoPhC.182.1384A})}

\appendix 
\section{SEDs of selected HAWC sources}\label{Sec:Appendix}
\begin{figure*}[hbtp!]
  \centering
  \includegraphics[width=\textwidth]{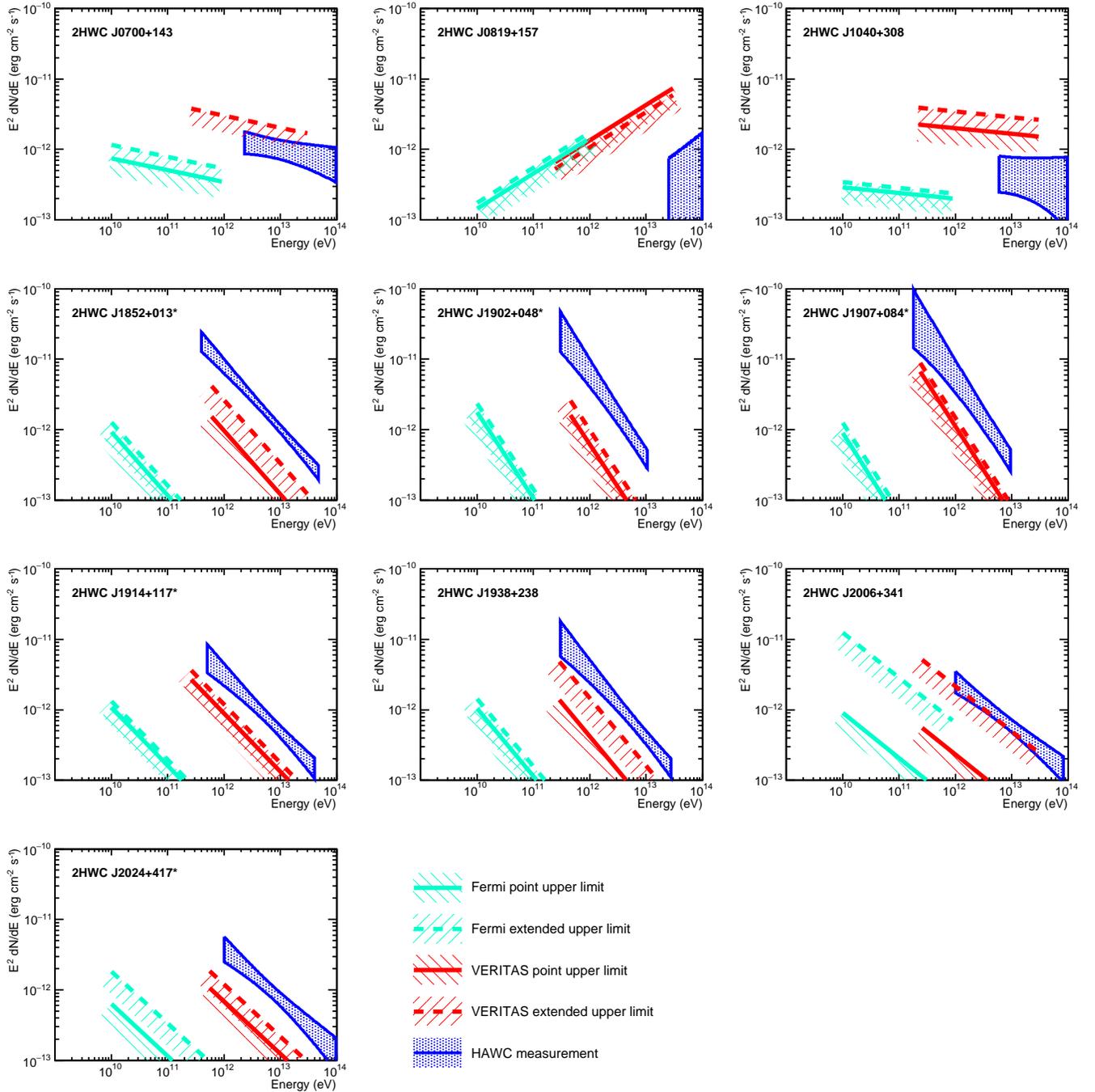}
\caption{SEDs of undetected HAWC sources. Upper limits for \textit{Fermi}-LAT(teal lines) and VERITAS(red lines) were calculated by using the spectral index estimated by HAWC.
\label{Fig:SEDs_undetected_sources_w_indexH}} 
\end{figure*}

\begin{figure*}[hbtp!]
  \centering
  \includegraphics[width=\textwidth]{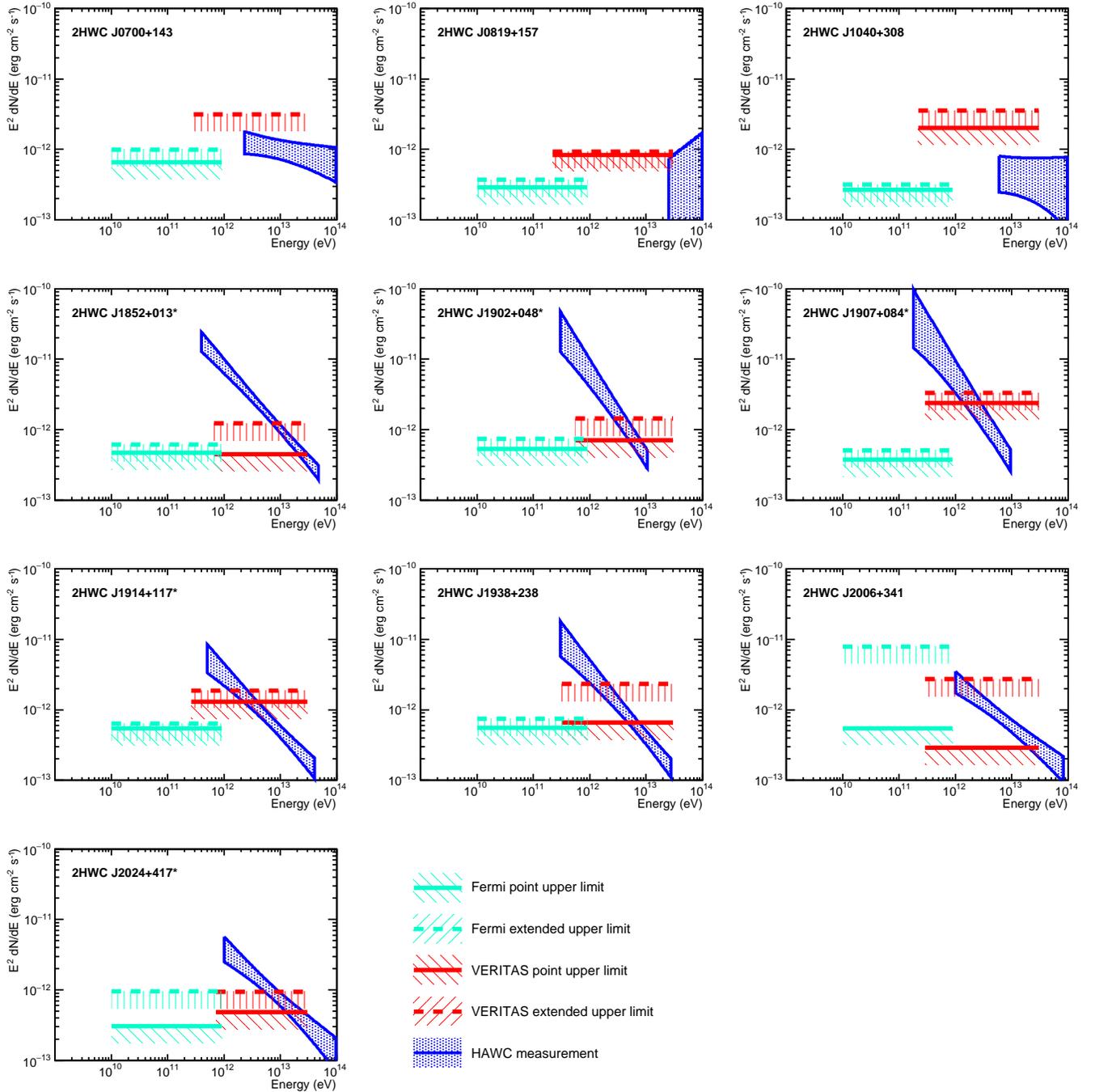}
\caption{SEDs of undetected HAWC sources. Upper limits for \textit{Fermi}-LAT(teal lines) and VERITAS(red lines) were calculated by using the spectral index of 2.} 
\label{Fig:SEDs_undetected_sources_w_index2} 
\end{figure*}

Figure~\ref{Fig:SEDs_undetected_sources_w_indexH} and Figure~\ref{Fig:SEDs_undetected_sources_w_index2} show individual SEDs of ten 2HWC sources out of fourteen selected 2HWC sources that were not shown in the previous sections. Upper limits from \textit{Fermi}-LAT and VERITAS with a point-like source assumption and with an extended source assumption are shown together with HAWC's flux measurements. For extended source studies, \textit{Fermi}-LAT assumed the size of the source to vary from a radius of 0.23${^\circ}$ to 1.0${^\circ}$ while VERITAS assumed a source radius of 0.23${^\circ}$. The details of the analyses and results can be found in Section~\ref{Sec:Analysis} and Table~\ref{Tbl:NonDetections}. In Figure~\ref{Fig:SEDs_undetected_sources_w_indexH}, the spectral index measured by HAWC was used to calculate the upper limits while a spectral index of 2 was used for Figure~\ref{Fig:SEDs_undetected_sources_w_index2}.

\bibliographystyle{aasjournal}
\bibliography{FVH.bib}
\end{document}